\documentclass[twocolumn,showpacs,preprintnumbers,amsmath,amssymb]{revtex4}
\usepackage{graphicx}
\usepackage{dcolumn}
\usepackage{bm}
\usepackage{color}

\newcommand{\kbar}{$\bar{K}$}

\newcommand{\KKNN}{$\bar{K}\bar{K}NN$}
\newcommand {\be}{\begin{equation}}
 \newcommand {\ee}{\end{equation}}
 \newcommand {\bea}{\begin{eqnarray}}
 
 \newcommand {\eea}{\end{eqnarray}}

\begin{document}
 
\markboth{}{$K^-K^-pp$}

\title{Strong binding and shrinkage\\
 of single and double \kbar~nuclear systems ($K^-pp$, $K^-ppn$, $K^-K^-p$ and $K^-K^-pp$)\\
  predicted by Faddeev-Yakubovsky calculations
 }


\author{\sc Shuji Maeda$^1$, Yoshinori {\sc Akaishi}$^{2,3}$ and Toshimitsu {\sc Yamazaki}$^{2,4}$}

\address{$^{1}$ Department of Agro-Environmental Science, Obihiro University of Agriculture and Veterinary Medicine, Obihiro, Hokkaido 080-8555, Japan}
\address{$^{2}$ RIKEN Nishina Center, Wako, Saitama 351-0198, Japan}
\address{$^{3}$ College of Science and Technology, Nihon University, Funabashi, Chiba 274-8501, Japan}
\address{$^{4}$ Department of Physics, University of Tokyo, Hongo, Bunkyo-ku, Tokyo 113-0033, Japan}

\date{Accepted, Sept. 6, 2013, for Proc. Japan Academy, Series B, Vol. {\bf 89} (2013) 418-437}

\begin{abstract}
Non-relativistic Faddeev and Faddeev-Yakubovsky calculations were made for $K^{-}pp$, $K^{-}ppn$, $K^{-}K^{-}p$ and $K^{-}K^{-}pp$ kaonic nuclear clusters, where the quasi bound states were treated as bound states by employing real separable potential models for the $K^{-}$-$K^{-}$ and the $K^{-}$-nucleon interactions as well as for the nucleon-nucleon interaction. 
The binding energies and spatial shrinkages of these states, obtained for various values of the $\bar{K}N$ interaction, were found to increase rapidly with the $\bar{K}N$ interaction strength. Their behaviors are shown in a reference diagram, where possible changes by varying the $\bar{K}N$ interaction in the dense nuclear medium are given.
Using the $\Lambda(1405)$ ansatz with a PDG mass of 1405 MeV/$c^2$ for $K^-p$, the following ground-state binding energies together with the wave functions were obtained: 51.5 MeV ($K^{-}pp$), 69 MeV ($K^{-}ppn$), 30.4 MeV ($K^{-}K^{-}p$) and 93 MeV ($K^{-}K^{-}pp$), which are in good agreement with previous results of  variational calculation based on the Akaishi-Yamazaki coupled-channel potential. 
The $K^{-}K^{-}pp$ state has a significantly increased density where the two nucleons are located very close to each other, in spite of the inner $NN$ repulsion. Relativistic corrections on the calculated non-relativistic results indicate substantial lowering of the bound-state masses, especially of $K^-K^-pp$, toward the kaon condensation regime. 
The fact that the recently observed binding energy of $K^-pp$ is much larger (by a factor of 2) than the originally predicted one may infer an enhancement of the $\bar{K}N$ interaction in dense nuclei by about 25\%, possibly due to chiral symmetry restoration. In this respect some qualitative accounts are given based on "clearing QCD vacuum" model of Brown, Kubodera and Rho. 

\end{abstract}

\maketitle     
   
\section{Introduction}

For the past decade we have studied deeply bound and dense kaonic nuclear cluster (KNC) states using an empirically based coupled-channel $\bar{K}N$ complex potential \cite{Akaishi02,Yamazaki02,Dote04a,Dote04b,Yamazaki04,Yamazaki07a,Yamazaki07b}, which is characterized by a strongly attractive $I=0~\bar{K}N$ interaction coupled with the $\Sigma \pi$ channels. 
Its strength was adjusted to reproduce the mass 1405 MeV/$c^2$ for the so-called $\Lambda(1405)$, which was assumed as the quasi-bound $K^-$-$p$ state (called $\Lambda^* \equiv \Lambda(1405)$ ansatz). The most spectacular prediction, first shown in \cite{Akaishi02}, is that light nuclei involving a $\bar{K}$ as a constituent are shrunk to dense baryonic objects with densities nearly 3-times as much as the normal nuclear density. The structure of the most basic system, $K^-pp$, first predicted in 2002 \cite{Yamazaki02}, was studied in detail by a realistic three-body calculation \cite{Yamazaki07a,Yamazaki07b}, and hence a molecular covalent nature of the strong binding was revealed. The $K^-pp$ system was shown to have a structure of $\Lambda^*$-$p$, where the quasi-bound $I=0$ $\bar{K}N$ pair that is identified as $\Lambda^* \equiv \Lambda(1405)$ behaves like an ``atom". This study led us to a new concept of nuclear force, ``super-strong nuclear force", which is caused by a real $\bar{K}$ migrating between two nucleons, conceptually similar to the covalent bonding in hydrogen molecules: H$_2^+$ and H$_2$ \cite{Heitler}, and has a binding strength nearly 4-times as large as that of the ordinary nuclear force. The uniqueness of this bonding is that it is caused by a strongly interacting real boson, $\bar{K}$. This new type of force was given a name by Nishijima \cite{Nishijima08}: ``kaonic origin of nuclear force", which is contrasted to the ordinary ``pionic origin of the nuclear force". 

In these papers \cite{Yamazaki07a,Yamazaki07b} we have predicted a new formation mechanism of $K^-pp$ in the process of $pp \rightarrow K^+ + p + \Lambda^* \rightarrow K^+ + K^-pp$, whereas a normal $\Lambda$ particle is produced in $pp \rightarrow K^+ + p + \Lambda$. The $pp$ collision at high energy is known to produce $\Lambda(1405)$ among other hyperons, as revealed in missing-mass spectra, $MM(p K^+)$ \cite{MM,Zychor}. It was theoretically clarified that the produced $\Lambda^*$ serves as a doorway to form a complex $\Lambda^*$-$p \approx K^-pp$. This sticking process occurs strongly when the formed $K^-pp$ has a small $p$-$p$ distance that matches the small proximity of the $pp$ collision with a short collision length of $\sim 0.3$ fm, helped by a large momentum transfer of $\sim$ 1.6 GeV/$c$ at $T_p \sim 3$ GeV. In other words, the doorway particle, $\Lambda^*$, sticks to the participating $p$ at an enormously high probability to form a $K^-pp$, if and only if it is dense, and thus the occurrence of this reaction with a large $K^-pp$ cross section would provide definite evidence for a dense $\bar{K}$ nuclear state. Thus, the $pp$ reaction has a great advantage for KNC production compared to statistical coalescence processes where deeply bound and dense states are hardly formed \cite{Cho11}.

Recently, this $pp$ reaction process was searched for in old DISTO experimental data of exclusive events of $pp \rightarrow p + \Lambda + K^+$ at $T_p = 2.85$ GeV, and the process was indeed found to take place \cite{Yamazaki10}. (An indication of $K^-pp$ had been reported before from a stopped-$K^-$ experiment on light nuclei with a similar binding energy \cite{FINUDA}.) A broad peak showing $M = 2267 \pm 2 (stat) \pm 5 (syst)$ MeV/$c^2$ was revealed with as much intensity as the free $\Lambda^*$ emission. The $\Lambda^*$ production observed at large angles of proton emission in $pp \rightarrow p + K^+ + \Lambda^*$ reactions \cite{Kienle12,HADES12} indicates that the $\Lambda^*$ production takes place in a short collision length of $pp$ \cite{Hassanvand13}, justifying the basis of our proposal. 
Thus, this experiment has indicated that the state observed is a dense $K^-pp$. In addition, the same reaction was studied at $T_p =$ 2.5 GeV, a substantially lower energy compared with 2.85 GeV. The same $K^-pp$ with $M = 2267$ MeV/$c^2$ was expected kinematically, but it was not populated at 2.5 GeV \cite{Kienle12,Suzuki13}. We have reached the following conclusion. At this lowered incident energy the $\Lambda^*$ production was very much decreased toward the production threshold, so that the production of $K^-pp$ through $\Lambda^*$ sticking was also suppressed. In this way, the role of $\Lambda^*$ as a doorway to form $K^-pp$ was understood.
The observed binding energy of $K^-pp$ in DISTO is much larger than the original prediction of \cite{Yamazaki04,Yamazaki07b}. Indeed, a view with a 25\% enhancement over the original $\bar KN$ interaction seems to be compatible with the DISTO observation. We call this enhanced empirical interaction "DISTO". This fact strongly suggests that we have to consider seriously possible enhancement effects of the $\bar KN$ interaction in dense KNC systems. 

The simplest double $\bar{K}$ nuclei, $K^-K^-pp$ and $K^-K^-ppn$, were also predicted to be deeply bound with binding energies of 117 MeV and 221 MeV, respectively \cite{Yamazaki04}. Extending the predicted and proven mechanism of a large sticking of $\Lambda^*$ to $p$, we proposed a similar formation mechanism for $K^-K^-pp$ via a double $\Lambda^*$-$\Lambda^*$ doorway \cite{Yamazaki11,Hassanvand11}. The $K^-K^-pp$ is expected to be formed abundantly by $pp$ collisions of an incident proton energy of about 7 GeV. Thus, the double $\bar{K}$ nuclei are within the reach of experimental studies in the near future at J-PARC and FAIR. It is extremely interesting to search for double-$\bar{K}$ nuclear clusters, since they may serve as precursors to kaon condensation \cite{Nelson,Brown,Brown-Bethe}. In view of the DISTO data, the $\bar{K}N$ interaction, and thus the actual binding of $K^-K^-pp$, may be much stronger than in the original prediction.

On the other hand, there are serious theoretical discrepancies concerning KNC bindings, essentially coming from the different starting ansatzes, whether A) the $\Lambda(1405)$ mass is at the traditional "phenomenological" value, 1405 MeV/$c^2$ \cite{PDG,Esmaili10,Esmaili11}, or B) a "chiral weak" value, 1420 MeV/$c^2$ or more together with a double-pole structure (for instance, see \cite{Hyodo08}). The latter {\it Chiral} ansatz leads to a substantially shallow $\bar{K}N$ potential, and thus to kaonic states of small binding energies \cite{Dote09}. Very recently, Barnea {\it et al.} \cite{Barnea12} made a hyperspherical harmonics calculation for the $K^-pp$, $K^-ppn$ and $K^-K^-pp$ systems 
based on the ``chiral weak'' interaction, and  
 obtained very shallow bound states. Naturally, all of these binding energies are substantially smaller than our earlier predictions, based on the gstrong regime" $\bar KN$ interaction \cite{Akaishi02,Yamazaki02,Dote04a,Dote04b,Yamazaki04,Yamazaki07a,Yamazaki07b}. We point out that there is no clear experimental evidence to support the ``chiral weak" ansatz and predictions. Recent HADES data on $pp \rightarrow p + \Lambda^* + K^+$ \cite{HADES12} is also in favour of the strong regime with the PGD value \cite{Hassanvand13}.  

In any case, we herewith introduce such a practical way as a ``reference diagram'' (explained below) to evaluate the difference of binding energies and sizes of KNC for various $\bar KN$ interactions for comparing different theories on equal footing. A possible change of the 
$\bar KN$ interaction due to the shrinkage of KNC is studied from the view point of chiral symmetry restoration. Clear evidence for partial restoration of chiral symmetry in a nuclear medium has been obtained from deeply bound pionic states \cite{Kienle04,Suzuki04,Yamazaki:PhysRep}, which indicates that the quark condensate that exists in the free $\pi N$ interaction decreases by $\sim 30$\% at the normal nuclear density. A similar and much more dramatic effect is expected in the case of $\bar{K}$ nuclear clusters because of the attractive character of the interaction, and thus of a tremendously amplified nuclear shrinkage effect, which we discuss at the end of this paper.

Since the developments of the Faddeev \cite{Faddeev:60} and the Faddeev-Yakubovsky \cite{Faddeev-Yakub:67} methods to solve the 3- and the 4-body Schr\"odinger equations, respectively, in the form of  integral equations, they have been successfully applied in many regions of few body physics because of their nature to give rigorous wave functions under assumed two-body interactions. 
This is due to the appropriate division of the 3-body (4-body) $t$-matrix and accordingly of the wave function into the Faddeev (Faddeev-Yakubovsky) components, so that the rigorous derivation of the wave function is most straightforwardly achieved under separable 2-body potentials. It has also been clearly shown that the Faddeev-Yakubovsky calculation can have a high precision under local realistic potentials, as shown by
 a benchmark test calculation of a four-nucleon bound state \cite{Kamada:01}. Application of this method has been extended to the strangeness nuclear physics by Filikhin and Gal \cite{Filikhin-Gal:02}. In the problem of $\bar{K}$-bound high-density nuclear systems, the Faddeev and the Faddeev-Yakubovsky methods should have an advantage to clarify the structure of the wave function in the central region of the system where more than two particles are very close to each other so that the wave function can be complicated, because, in contrast to the variational methods and the wave functional expansion methods like the hyperspherical harmonics method, they presuppose no structures of the wave function.

$K^{-}pp$ has been treated by Shevchenko, Gal and Mares \cite{Shevchenko07a}, 
Shevchenko, Gal, Mares, and R\'evai  \cite{Shevchenko07b}, and Ikeda and Sato \cite{Ikeda07,Ikeda09,Ikeda10} in the integral equation formalism. In these calculations, they used the Alt-Grassberger-Sandhas (AGS) equation \cite{Alt67} to find the resonance energies and widths. The AGS equation is directly connected to the scattering matrix elements via the channel resolvents, and is thus fit to calculate reaction quantities, but is not directly connected to the wave function. This is the reason why these authors derived only the energies and widths of $K^{-}pp$. On the other hand, to study the shrinkage and formation of dense nuclear configurations inside the $KNC$ systems induced by the strongly attractive $\bar{K}N$ interaction, deriving the wave functions of the $KNC$ states is indispensable. In nuclear bound states it is a general feature that the bound-state energy arises as a strong cancellation between the kinetic energy and the potential energy, and since the dense structure of the system may be linked to a large kinetic energy expectation value, the calculation of this quantity from the wave function may contain important information about the shrinkage behavior of the $KNC$ systems. In the present Faddeev and Faddeev-Yakubovsky calculations, we put emphasis on the derivation and detailed analysis of the wave functions.

\section{Faddeev-Yakubovsky Formalism}

For systematic calculations of the ground-state energies and the density distributions of the light $\bar{K}$-nuclear systems, we performed 3-body and 4-body calculations with separable pair potentials in the non-relativistic Faddeev and Faddeev-Yakubovsky formalisms. 
In the Faddeev-Yakubovsky formalism, the bound-state wave function of a four-particle system $\Psi$ is written as a sum of two kinds of components (Faddeev-Yakubovsky components) $\psi$ and $\chi$,
\be
\Psi = \sum_{\alpha}^{12}\psi(\vec{k}_\alpha,\vec{p}_\alpha,\vec{q}_\alpha)
 +  \sum_{\beta}^{6}\chi(\vec{k}_\beta,\vec{\kappa}_\beta,\vec{s}_\beta),
\ee
where $(\vec{k}_\alpha,\vec{p}_\alpha,\vec{q}_\alpha)$ and 
$(\vec{k}_\beta,\vec{\kappa}_\beta,\vec{s}_\beta)$ 
 are two kinds of sets of Jacobi momenta for the four-particle system, and $\alpha$ and $\beta$ correspond to different choices of these sets of Jacobi momenta.
 It is well known that the Faddeev and the Faddeev-Yakubovsky components are rather simple in structure compared with the total wave functions. This would be a great advantage of the methods for such a calculation as the present one, where the different natures of the different pair interactions together with the strong binding would bring complex structures into the wave functions. 
 
 In the case of a separable S-wave two-body potential,
\be 
v(\vec{k},\vec{k'}) = \sum_{i}\lambda_{i}g_{i}(k)g_{i}(k')\frac{1}{4\pi},
\ee	 
the $t$-matrix is also an S-wave, and is written as
\bea
&&t(\vec{k},\vec{k'};z) = \sum_{ij}g_{i}(\textit{k})\tau_{ij}(z)g_{j}(\textit{k}')\frac{1}{4\pi} \nonumber \\
&&[\tau^{-1}(z)]_{ij} = \frac{1}{\lambda_{i}}\delta_{ij} - \int_{0}^{\infty}k^{2}dk\frac{g_{i}(k)g_{j}(k)}{z - \frac{\hbar^{2}}{2\mu}k^{2}},
\eea
where $z$ is the two-body energy and $\mu$ is the reduced mass of the 2-body system. In this case, 
the Faddeev-Yakubovsky components are written in the form
\bea
&&\psi^{\alpha}(\vec{k}_\alpha,\vec{p}_\alpha,\vec{q}_\alpha)
= \frac{1}{z - T^{\alpha}(k_{\alpha},p_{\alpha},q_{\alpha})} \nonumber \nonumber \\
&&\times \sum_{ij}
g_{i}^{\alpha}(k_{\alpha})\tau_{ij}^{\alpha}(z-T^{\alpha}(p_{\alpha},q_{\alpha}))
A_{j}^{\alpha}(\vec{p}_{\alpha},\vec{q}_{\alpha};z)\\
&&\chi^{\beta}(\vec{k}_\beta,\vec{\kappa}_\beta,\vec{s}_\beta)
 = \frac{1}{z - T^{\beta}(k_{\beta},\kappa_{\beta},s_{\beta})} \nonumber \\
&&\times  \sum_{ij}
g_{i}^{\beta}(k_{\beta})\tau_{ij}^{\beta}(z-T^{\beta}(\kappa_{\beta},s_{\beta}))
B_{j}^{\beta}(\vec{\kappa}_{\beta},\vec{s}_{\beta};z),
\eea
where $z$ is the four-body energy, $T^{\alpha}(k_{\alpha},p_{\alpha},q_{\alpha})$ and $T^{\beta}(k_{\beta},\kappa_{\beta},s_{\beta})$ are the four-body kinetic energies represented by the Jacobi momenta, and $T^{\alpha}(p_{\alpha},q_{\alpha})$ and $T^{\beta}(\kappa_{\beta},s_{\beta})$ are the spectator kinetic energies. For a 2+2 system, there are 4 different functions, $\psi^{\alpha}$, and 3 different functions, $\chi^{\beta}$, while for a 3+1 system, there are 3 different functions, $\psi^{\alpha}$, and 2 different functions, $\chi^{\beta}$. $A^{\alpha}_{j}$ and $B^{\beta}_{j}$ are solved by a set of coupled integral equations: 
\bea
&&A_{j}^{\alpha}(\vec{p}_{\alpha},\vec{q}_{\alpha};z)
 = \sum_{j'j''\alpha'}\int d\vec{q}_{\alpha'} \nonumber \\
&&\times X_{jj'}^{\alpha\alpha'}(\vec{p}_{\alpha},\vec{p}_{\alpha'};z-T^{\alpha}(q_{\alpha})
)\tau_{j'j''}^{\alpha'}(z-T^{\alpha'}(p_{\alpha'},q_{\alpha'})) \nonumber \\
&&\times A_{j''}^{\alpha'}(\vec{p}_{\alpha'},\vec{q}_{\alpha'};z) \nonumber \\
&& + \sum_{j'j''\beta'}\int d\vec{s}_{\beta'} \nonumber \\
&&\times X_{jj'}^{\alpha\beta'}(\vec{p}_{\alpha},\vec{\kappa}_{\beta'};z-T^{\alpha}(q_{\alpha}))
\tau_{j'j''}^{\beta'}(z-T^{\beta'}(\kappa_{\beta'},s_{\beta'})) \nonumber \\
&&\times B_{j''}^{\beta'}(\vec{\kappa}_{\beta'},\vec{s}_{\beta'};z) \\
&&B_{j}^{\beta}(\vec{\kappa}_{\beta},\vec{s}_{\beta};z)
 = \sum_{j'j''\alpha'}\int d\vec{q}_{\alpha'} \nonumber \\
&&\times Y_{jj'}^{\beta\alpha'}(\vec{\kappa}_{\beta},\vec{p}_{\alpha'};z-T^{\beta}(s_{\beta}))
\tau_{j'j''}^{\alpha'}(z-T^{\alpha'}(p_{\alpha'},q_{\alpha'})) \nonumber \\
&&\times A_{j''}^{\alpha'}(\vec{p}_{\alpha'},\vec{q}_{\alpha'};z), 
\eea
where $X_{jj'}^{\alpha\alpha'}$ and $X_{jj'}^{\alpha\beta'}$ are the [3+1] subsystem amplitudes and 
$Y_{jj'}^{\beta\alpha'}$ are the [2+2] subsystem amplitudes, with $T^{\alpha}(q_{\alpha})$ and $T^{\beta}(s_{\beta})$ the spectator kinetic energies with respect to these subsystems.

 We make a separable representation for these 
subsystem amplitudes by the energy dependent pole expansion (EDPE) 
\cite{Sofianos79}. The convergence behavior of this expansion has been investigated in the bound states of the cluster of 4 helium atoms \cite{Nakaichi82}, and found to be fast compared with that of the Hilbert-Schmidt expansion \cite{Narodetsky}, so that retaining only one term for each subsystem amplitude already achieves a good approximation. The system of 4 helium atoms was also treated by the Faddeev-Yakubovsky equation in configuration space by Filikhin {\it et al.} within the cluster expansion method \cite{Filikhin02}, and their value agreed with ours by the first two digits. In the present study, we retained only 1 term of EDPE for each subsystem amplitude, {\it i.e.}, the amplitude dominated by the subsystem ground state pole (EDPA). 

For the 2-body potentials, we adopted S-wave separable potentials with Yamaguchi form factors:
\bea
&&\lambda_i = \frac{4}{\pi} \frac{\hbar^2}{2 \mu} \frac{s_i}{\beta_i}, \nonumber \nonumber \\
&&g_i (k) = \frac{\beta_i ^2}{k^2 + \beta_i ^2},
\eea	 
where $\mu$ is the reduced mass of the 2-body system. For the $NN$ potential we used the rank-2 potentials of Ikeda and Sato \cite{Ikeda07}:
\bea
^{1}S_{0}(I=1) : &&s_{R} = 7.40, \beta_{R} = 6.157 ~{\rm fm}^{-1}, \nonumber \\
&&s_{A} = -2.48, \beta_{A} = 1.784 ~{\rm fm}^{-1}, \nonumber \\
^{3}S_{1}(I=0) : &&s_{R} = 7.40, \beta_{R} = 6.157 ~{\rm fm}^{-1}, \nonumber \\
&&s_{A} = -3.26, \beta_{A} = 1.784 ~{\rm fm}^{-1},
\eea
where suffixes $R$ and $A$ indicate the repulsive and attractive terms, respectively.

For the $\bar{K}N$ and $\bar{K}\bar{K}$ potentials we used rank-1 potentials:
\bea
\bar{K}N : &&s_{(I=0)} = -1.37, \beta_{(I=0)} = 3.9 ~{\rm fm}^{-1}, \nonumber \\
&&s_{(I=1)} = 0.29 \cdot s_{(I=0)}, \beta_{(I=1)} = \beta_{(I=0)},  \nonumber \\
\bar{K}\bar{K} : &&s_{(I=1)} = 0.38, \beta_{(I=1)} = 3.9 ~{\rm fm}^{-1}.
\eea
We call these short-ranged potentials 
the AMY (Akaishi-Myint-Yamazaki \cite{Akaishi08}) $\bar{K}N$ and $\bar{K}\bar{K}$ potentials. The range parameter values $3.9~ {\rm fm}^{-1}$ are adopted to represent the exchange of heavy mesons, like the $\rho$ meson.

Among the \kbar ~nuclear systems studied in the present work, the $K^-K^-pp$ system is particularly an interesting system because it contains various features of interaction: a weakly repulsive interaction between the two anti-kaons, an interaction between the two nucleons with long-range attraction and short-range repulsion, and a short-range strong attraction between an \kbar ~and a nucleon. We found that, among these, the $I = 0$ $\bar{K}N$ attraction plays a dominant role to determine the structure of the ground-state wave function.

The $\bar{K}N$ pair assumes two isospin states: $I=0$ and $I=1$. To incorporate these multi-state isospin configurations, we made a simplified treatment; we calculated the $t$-matrix in the $I=0$ state with the $I=0$ potential and the $t$-matrix in the $I=1$ state with the $I=1$ potential, and averaged these $t$-matrices with the weights $w=0.75$ and $1-w=0.25$, respectively, for the $K^-K^-p$, $K^-pp$, $K^-K^-pp$ systems, and $w=0.5$ and $1-w=0.5$ for the $K^-ppn$ system. It is expected that this treatment (``$t$-averaging'' calculation) takes into account the strongly attractive $I=0$ interaction more properly than averaging the $I=0$ and $I=1$ potentials and obtaining a $t$-matrix for the $\bar{K}N$ system by the averaged $\bar{K}N$ potential ($V$-averaging calculation). We followed a similar prescription for the $NN$ configurations in the $K^-ppn$ calculation. The $t$-matrix averaging Faddeev and Faddeev-Yakubovsky calculations have an advantage for systematic calculations, like the present study, because it is an easy way to incorporate multi-state pair interactions when the interaction in one state is very strong and the interaction in the other states is weak. This effect will be demonstrated below by comparisons between the results for the bound state energies obtained by the $t$-matrix averaging and by the potential averaging calculations.

 From the obtained wave functions we calculated the density distributions of the pair distances and of the distances of the anti-kaons and the nucleons from the center of masses, in the $K^-K^-p$, $K^-pp$, $K^-K^-pp$ and $K^-ppn$ systems. We show the formula for the density distributions $\rho (r)$ of the pair distances in the 3-body $L=0$ states, where $L$ is the total orbital angular momentum, in eq. (\ref{eq:II.11}). The formulas for the distance of particles from the c.m. of the hypernuclei and formulas for the 4-body $L=0$ states were derived by a straightforward extension of this formula. 
\bea 
1&=&<\psi\mid\psi> \nonumber\\
  &=&\frac{2}{\pi}\int_{0}^{\infty}r^{2}dr\int d\vec{k}\int d\vec{k'}\int d\vec{p}\nonumber\\
 &\times& \psi(\vec{k}\vec{p})\psi(\vec{k'}\vec{p})\sum_{\lambda}\frac{2\lambda+1}{4\pi}j_{\lambda}(kr)j_{\lambda}(k'r)P_{\lambda}(\hat{k}\cdot\hat{k'}) \nonumber \\
&\equiv& \int_{0}^{\infty}r^{2}dr\rho(r).
\label{eq:II.11}\eea
Here, $\psi$ is the normalized 3-body wave function, which is expressed as a sum of 3 Faddeev components, while $\vec{k}$ and $\vec{p}$ are the Jacobi momenta. In the case of the 4-body systems, the wave functions are represented as a sum of 18 Faddeev-Yakubovsky components given by three Jacobi momenta. The multiple integration was made in the Monte Carlo scheme.

The density distributions are expressed as a sum of the $\lambda$ components, where $\lambda$ is the orbital angular momenta of the pairs or of a particle relative to the c.m. of the remaining part of the KNC system. For the ground states of the systems studied in the present work, it is found that the components with $\lambda > 0$ are negligible, so we retain only the terms with $\lambda=0$ in the present study.

\section{Results and discussion}

\subsection{Comparison between $t$-averaging and $V$-averaging procedures}

First, let us see the ground state energies of the $K^-p$, $K^-K^-p$, $K^-pp$, $K^-ppn$, and $K^-K^-pp$ systems, as listed in Table I, where the results obtained by the $t$-averaging calculations are compared with those by the $V$-averaging calculations.

\begin{table}[h]
\caption{\label{tab:Table1}  Comparison of the calculated ground-state energies between the $t$-averaging and $V$-averaging procedures. }
\begin{center}
\begin{tabular}{lrr}
\hline
State   ~~~& ~~~$t$-averaging   & ~~~~$V$-averaging  \\
\hline
$K^-p~(I=0)$              & -26.6 MeV        &  -26.6 MeV   \\
$K^-K^-p~(I=1/2)$   & -30.4 MeV         &  -5.0 MeV \\ 
$K^-pp~(I=1/2)$        & -51.5 MeV         &  -23.6 MeV \\
$K^-ppn~(I=0)$          & -69 MeV           & -31 MeV  \\
$K^-K^-pp~(I=0)$       & -93 MeV           & -54 MeV\\
\hline
\end{tabular}
\end{center}
\end{table}

  Here, we can observe that the ground states of the $K^-K^-p$, $K^-pp$, $K^-K^-pp$ and $K^-ppn$ systems gain appreciable binding when we take the $t$-averaging procedure rather than the $V$-averaging procedure. This indicates the fact that the strong $\bar{K}N$ attraction in the $I = 0$ state is well taken into account in the $t$-averaging calculations. As a result, we obtained $-51.5$ MeV for the ground-state energy of the $K^-pp$ system, which is more than a factor of 2 deeper than that we obtain in the $V$-averaging case. 
 
 In the early 1960's, just after the discovery of the $\Lambda(1405)$ resonance \cite{Alston61,Bastien61}, following the prediction of a $K^-p$ quasi-bound state by Dalitz and Tuan \cite{Dalitz59},  
  Nogami investigated the possible existence of $\bar{K}NN$ bound states, and claimed that $K^-pp$ may exist as a bound state with a binding energy of $B(K^-pp)=11.5$ MeV \cite{Nogami63}. Sometimes, this paper is referred to as the first prediction of the $K^-pp$ bound state \cite{Gal07,Bayar11}. 
   However, such citations are theoretically inadequate. It should be noticed that the $B(K^-pp)=11.5$ MeV is much less than $B(K^-p)=29.9$ MeV that Nogami reproduced with his 
   $\bar KN$ interaction; the $K^-pp$ comes above the $\Lambda(1405) + p$ threshold. Thus, Nogami's $K^-pp$ is not a bound state at all, and its escape width is estimated to be on the order of 300 MeV. Here, we have re-calculated the $K^-pp$ system with Nogami's $\bar KN$ interaction using the same $V$-averaging treatment, which yielded a $K^-pp$ bound state with $B(K^-pp)=60.3$ MeV. We have found that this $K^-pp$ lies much more deeply than his state with $B(K^-pp)=11.5$ MeV. (Since his potential is of long range (two-$\pi$ range), the $V$-averaging value is not so different from the $t$-averaging one.) 
   Nogami's paper fatally missed just this genuine $K^- pp$ ground state. After many years, the first prediction of $K^-pp$ quasi-bound state below the $\Lambda(1405)+p$ threshold was given in 2002 by Yamazaki and Akaishi \cite{Yamazaki02}. 

The first prediction of $K^-pp$ energy was given to be $-47.7-i30.6$ MeV based on a variational method of Akaishi \cite{Akaishi86} with the following strongly isospin dependent $\bar KN$ interactions;
\begin{eqnarray}
V^{I=0}_{\bar KN}(r)=(-595.0-i83.0)~{\rm exp}[-(r/0.66~{\rm fm})^2]~~{\rm MeV}, \nonumber \\ 
V^{I=1}_{\bar KN}(r)=(-175.0-i105.0)~{\rm exp}[-(r/0.66~{\rm fm})^2]~~{\rm MeV}. \nonumber\\
\end{eqnarray}
Here we show $V$-averaging and $t$-averaging results for this realistic $K^-pp$ case in order to know the accuracy of our isospin averaging procedures. The $V$-averaging treatment gives an energy of $-33.9 -i29.9$ MeV, where higher-order effects of the $V^{I=0}_{\bar KN}$ strong attraction are improperly suppressed by the averaging with the $V^{I=1}_{\bar KN}$ weaker attraction. Most of such improper suppression can be removed by calculating separately the $I=0$ and $I=1$ $t$-matrices which incorporate their higher-order effects respectively. By simulating the isospin average of both the $t$-matrices we have obtained the $t$-averaging energy of the $K^-pp$ to be $-45.4-i32.4$ MeV, which becomes much closer to the original value, $-47.7-i30.6$ MeV. Thus, we can estimate the error of the $t$-averaging to be $2 \sim 3$ MeV for realistic $K^-pp$. 

 As for the $K^-K^-p$ system, Kanada-En`yo and Jido \cite{Kanada08} made a variational calculation by a Gaussian expansion method. They obtained a bound state (with the ground state energy $-36.0$ MeV) slightly below the $K^-p$ threshold ($-30.6$ MeV), when the Akaishi $K^-p$ effective potential was employed. We obtained a similar result, i.e. $-30.4$ MeV for the ground-state energy of the $K^-K^-p$ system just below the $K^-p$ threshold $-26.6$ MeV. It is noted that we could not obtain such a bound state of the $K^-K^-p$ system by the $V$-averaging procedure.

Barnea {\it et al.} \cite{Barnea12} made a hyperspherical harmonics calculation for the $K^-pp$, $K^-K^-pp$ and $K^-ppn$ systems using the shallow chiral interaction model with the self-consistent energy dependence taken into account. They reproduced the ground-state energies of the $K^-p$ and $K^-pp$ systems by Dot$\acute{\rm e}$ {\it et al.} \cite{Dote09}, based on the shallow chiral model, and obtained the ground-state energies of the $K^-K^-pp$ and $K^-ppn$ systems around $-30$ MeV. On the other hand, our calculation, based on the gstrong regimeh $\bar{K}N$ interaction, leads to very deep ground-state energies, i.e., $-69$ MeV and $-93$ MeV for the $K^-ppn$ and $K^-K^-pp$ systems, respectively. In particular, it should be noted that the addition of one nucleon to the $K^-K^-p$ system gains $-62$ MeV, and the addition of one $\bar{K}$ to the $K^-pp$ system gains $-41$ MeV to the ground-state energy. Figure 1 shows a comparison of the present calculations with the result of a chiral-based theory of Barnea {\it et al.} [29] (black broken bars). Whereas in the latter calculation the $K^-pp$ through $K^-K^-pp$ levels are shallow and nearly flat (around -30MeV), our result with the standard strength of the $\bar{K}N$ interaction ($s_{\bar{K}N}^{(I=0)}$=-1.37) is characterized by a sudden drop of the $K^-K^-pp$ energy. This behavior is more significant with the increase of the attractive interaction toward $s_{\bar{K}N}^{(I=0)}$=-1.60.\begin{table*}
\caption{\label{tab:table2} Calculated $s_{\bar{K}N}^{(I=0)}$ dependences of the ground-state energies ($E$ in MeV), the nucleon {\it rms} distributions for a point nucleon ($R_{KNC}^{\rm point}$ in fm), the nucleon {\it rms} distributions for a finite-sized nucleon ($R_{KNC}$ in fm),  
  {\it rms} $N$-$N$ distances ($R_{NN}$ in fm) and the effective densities $\rho^{\rm eff}/\rho_0$  for $K^-p$, $K^-K^-p$, $K^-pp$, $K^-ppn$ and $K^-K^-pp$. The values for the standard strength of the $\bar{K}N$ interaction ($s_{\bar{K}N}^{(I=0)}$ = -1.37) are shown in gothic letters.} 

\begin{center}
{\footnotesize 
\begin{tabular}{l|ccc|ccc}
\hline \hline
 &\multicolumn{3}{c}{$K^-p$} &\multicolumn{3}{c}{$K^-K^-p$} \\
 \hline 
$s_{\bar{K}N}^{(I=0)}$  &  $E$  & $R_{KNC}^{\rm point}$  &  $R_{KNC}$ &  $E$  & $R_{KNC}^{\rm point}$ & $R_{KNC}$\\ \hline

-1.2  & -8.3 & 0.75 & 1.16 & -9.1 & 1.59 & 1.82    \\
-1.3  & -18.0 & 0.55  & 1.04&  -20.1 & 1.05 & 1.37   \\
{\bf -1.37} & {\bf-26.6} & {\bf 0.47} & {\bf 1.00} & {\bf -30.4} & {\bf 0.84} & {\bf 1.22}  \\
-1.4 & -30.7 & 0.44 & 0.98 &  -35.4 & 0.78 &  1.18    \\
-1.5  & -46.2 & 0.38 &  0.96&    -54.8 & 0.63 & 1.08  \\
-1.6 & -64.2 & 0.34 & 0.94 &  -78.1 & 0.54 & 1.03 \\
-1.7 & -84.4 & 0.31 & 0.93 & -105.1 & 0.47 &  1.00    \\
\hline \hline
\end{tabular}
}
\end{center}

{\footnotesize 
\begin{ruledtabular}
\begin{tabular}{l|ccccc|ccccc|ccccc}
 &\multicolumn{5}{c}{$K^-pp$} &  \multicolumn{5}{c}{$K^-ppn$} &  \multicolumn{5}{c}{$K^-K^-pp$} \\
 \hline
$s_{\bar{K}N}^{(I=0)}$  &   $E$  & $R_{KNC}^{\rm point}$ & $R_{KNC}$&  $R_{NN}$ & $\rho^{\rm eff}/\rho_0$ & $E$  & $R_{KNC}^{\rm point}$ & $R_{KNC}$& $R_{NN}$ & $\rho^{\rm eff}/\rho_0$ & $E$  & $R_{KNC}^{\rm point}$ & $R_{KNC}$ & $R_{NN}$ & $\rho^{\rm eff}/\rho_0$\\ \hline

-1.2  & -23.8 & 1.06 & 1.38 & 1.93 & 1.48 & -42 & 1.15 & 1.45 & 1.89  & 1.58& -43  &  0.91 & 1.27 &  1.57 & 2.75\\
-1.3   & -39.0 & 0.94 & 1.29 & 1.73 & 2.06 & -57 & 1.09 & 1.40 & 1.80  &1.83 & -70 & 0.82& 1.20& 1.43 & 3.64      \\
{\bf -1.37}   &{\bf -51.5}   & {\bf 0.89}  & {\bf 1.25}& {\bf 1.62}  &  {\bf 2.50} & {\bf -69}  & {\bf 1.06}  & {\bf 1.37} & {\bf 1.75}  & {\bf 1.99} & {\bf -93}   &  {\bf 0.76}  & {\bf 1.16} & {\bf 1.35} & {\bf 4.33}   \\
-1.4 &  -57.3   &    0.86  &  1.23 &  1.58 & 2.70 &  -74  & 1.05  & 1.37 & 1.73  & 2.06 & -104 &  0.74  & 1.15 & 1.31  & 4.73  \\
-1.5  &   -78.3   &   0.81   & 1.19 &  1.48 & 3.28 &    -95  &  1.00  & 1.33 & 1.66 & 2.33 &  -144  &  0.69  &  1.12 &  1.22 & 5.86   \\
-1.6 &    -101.9 &   0.76 &  1.16 &   1.40 &  3.88 &  -117  & 0.97  & 1.31 & 1.61  & 2.55 & -190  &   0.66  & 1.10 & 1.09 & 8.22   \\
-1.7 &  -127.9  &   0.72  & 1.14 &  1.32 & 4.63 &  -142 & 0.94  & 1.29 & 1.56 & 2.80 & -241  &  0.63   & 1.08&  0.94  & 12.8    \\
\end{tabular}
\end{ruledtabular} }
\end{table*}

\subsection{Overview of binding energies and sizes}

The analysis of the DISTO experiment on $pp \rightarrow p + K^+ + \Lambda$ reactions  showed a peak structure in the $K^+$  missing mass, $\Delta M(K^+)$, and $M(p \Lambda)$ invariant mass spectra, indicating a compact $K^-pp$ state formed with a binding energy of 103 MeV \cite{Yamazaki11}. The observed binding energy is much larger than the theoretical prediction based on the strong $\bar{K}N$ regime, and suggests that in dense nuclear matter, the $\bar{K}N$ interaction may become stronger than in free space. Thus, it is interesting to systematically investigate the binding of the $\bar{K}$-nuclear systems, when the $\bar{K}N$ interaction is changed. 

Since the structure of $\bar{K}$ nuclear states totally depends on the assumed $\bar{K}N$ interaction, we calculated the ground-state energies and {\it rms} nuclear radii as functions of the interaction strength, $s_{\bar{K}N}^{(I=0)}$. Figure 2 shows the behaviors of the ground state energies of the $K^{-}p$, $K^{-}K^{-}p$, $K^{-}pp$, $K^{-}ppn$ and $K^{-}K^{-}pp$ systems when $s_{\bar{K}N}^{(I=0)}$ is varied, while the ratio of $s_{\bar{K}N}^{(I=1)}$ to $s_{\bar{K}N}^{(I=0)}$ and the other potential parameters are fixed.

The interaction parameter that we adopt ranges from the weak chiral regime ($s_{\bar{K}N}^{(I=0)} \approx -1.2$), as given in many chiral theories \cite {Hyodo08,Dote09,Barnea12}, and a much stronger regime of our hypothetical concern ($s_{\bar{K}N}^{(I=0)} \approx -1.5, -1.6, -1.7$), which may be required to account for experiments. The standard "PDG" value with $M(\Lambda^*) = 1405$ MeV/$c^2$ corresponds to
\be
{\rm ``Standard"}:~ s_{\bar{K}N}^{(I=0)} ({\rm Std}) = -1.37.
\ee  
 In this way, we can overview any theoretical predictions and the present and future observations. In Table~\ref{tab:table2} and Fig.2 we summarize our results.

It has been well known since the publication of \cite{Akaishi02} that the strong attractive $\bar{K}N$ interaction makes the $\bar{K}$ bound system shrunk, despite the strong short-range $NN$ repulsion. Since our calculation is capable of providing the wave functions, we calculated the {\it rms} radii of the bound nucleons, $R_{KNC}$, of each $\bar{K}$ nuclear cluster ($KNC$) from the obtained nucleon wave function as 
\be
R_{KNC}^{\rm point} = \sqrt{\frac{<\sum_{i}^{n_{N}}(\vec{r}_{i}-\vec{r}_{G})^{2}>}{n_{N}}},
\ee  
where $n_{N}$ is the number of the nucleons in the $KNC$ system, $\vec{r}_{i}$ is the position of the i-th nucleon, $\vec{r}_{G}$ is the position of the center of mass of the $KNC$  system, and the bracket $<>$ denotes an expectation value with a  normalized wave function of $KNC$. In the definition of $R_{KNC}^{\rm point}$, each nucleon is assumed to be a point. Then, we convoluted the nucleon size from the known proton radius \cite{PDG},
\be
r_p^{rms} = 0.88~{\rm fm},
\ee
into $R_{KNC}$ to obtain an {\it rms} nuclear radius of
\be
R_{KNC} = \sqrt{(R_{KNC}^{\rm point})^2 + (r_p^{rms})^2}.
\ee 
The results are also listed in Table~\ref{tab:table2}.

Table~\ref{tab:table2} presents an overview of calculated energies and nuclear radii of various $\bar{K}$ nuclear clusters at any input $\bar{K}N$ interaction strength. We can notice the following characteristics: \\

i) The binding energy increases with $|s_{\bar{K}N}^{(I=0)}|$. The degree of the increase is nearly the same for single-$\bar{K}$ clusters, whereas the degree of the binding energy increase for $K^-K^-pp$ is almost twice as large as those of the single-$\bar{K}$ clusters. This large dependence of the $K^-K^-pp$ energy on $|s_{\bar{K}N}^{(I=0)}|$ is understood from the different roles played by the most important potential energy, $V_{\bar{K}N}^{(I=0)}$, whose weight in the energy is 1 for $\bar{K}N$, 3/2 for $\bar{K}\bar{K}N$, $\bar{K}NN$ and $\bar{K}NNN$, and 3 for $\bar{K}\bar{K}NN$. \\

ii) The nuclear radii by nucleon distributions ($\bar{K}$ excluded) decrease with $|s_{\bar{K}N}^{(I=0)}|$, as initially predicted from the variational calculations of \cite{Akaishi02}. The nuclear shrinkage due to the attractive $\bar{K}N$ interaction is shown here as a general feature of $\bar{K}$ nuclear systems. The very much shrunk $K^-K^-pp$ system is of particular importance. However, the behavior of the nuclear sizes with $|s_{\bar{K}N}^{(I=0)}|$ is different from that of the binding energies. The nuclear density seems to be saturated, where the binding energy still increases. \\

\subsection{Higher-order effects of increased $\bar KN$ attractions}

When the $\bar KN$ attraction is increased, the potential, kinetic and binding energies of kaonic nuclear clusters are drastically enlarged, as shown in Fig. 3 (a) for $K^-p$, (b) for $K^-pp$ and (c) for $K^-K^-pp$.

In the case of the $K^-p$ system, (a), only a 15\% increase of the $K^-p$ interaction makes the binding energy of the system 2.24-times as large as that for the standard strength. Its reason can be understood as follows. At the standard strength, $s^{(I=0)}_{\bar KN}({\rm Std})=-1.37$, the binding energy of the $K^-p$ system, 26.6 MeV, is a result of large cancellation between the potential energy of -182.9 MeV and the kinetic energy of 156.2 MeV. At the 15\% strengthened $s^{(I=0)}_{\bar KN}$ interaction the potential energy of the system changes to -293.6 MeV with a 61\% enhancement in the absolute value, where 46(=61-15)\% comes from the higher-order effect of the increased $\bar KN$ attraction, which gives rise to a shrinkage of the wave function. This shrinkage also enhances the kinetic energy from 156.2 MeV to 233.9 MeV, partly cancelling the enhancement of the potential energy. In sum, the binding energy of $K^-p$ increases more than double from 26.6 MeV to 59.7 MeV due to the 15\% increase of the super-strong $K^-p$ attractive interaction.

Similar enhancement behaviors are seen for $K^-pp$, (b) and for $K^-K^-pp$, (c). In the case of $K^-pp$ the interaction between $K^-$ and the $pp$ core, which can be derived by a convolution procedure, becomes of longer range than the original $K^-p$ interaction. Generally, the higher-order effect of long-range interaction becomes less effective so that in limit the potential energy of the system is well-estimated in a Born approximation. This long-range property of the $K^-pp$ interaction explains that the enhancement factors in the $K^-pp$ case are rather damp compared with those in the $K^-p$ case. In the case of $K^-K^-pp$, where the number of $K^-p$ interaction is multiplied, the enhancement factors again rapidly increase toward quite a large binding as seen in Fig. 3 (c). Experimental information is awaited especially for highly compact system, $K^-K^-pp$.

\subsection{Density distributions of $\bar{K}$ nuclear clusters}

To investigate the feature of these large bindings and compact sizes, we analyzed the wave functions obtained in the present calculation. We show here some remarkable consequences.

\subsubsection{$N$-density distributions}

 Figure 4 (left) shows the nucleon density distributions, $\rho_{N}(r)$, where $r$ is the distance of the nucleon from the center of mass of (a) $K^{-}pp$, (b) $K^{-}ppn$ and (c) $K^{-}K^{-}pp$ 
for various values of $s_{\bar{K}N}^{(I=0)}$  = -1.2, -1.3, -1.4, -1.5, -1.6 and -1.7.
The nucleon density distribution in $K^-pp$ has a depression near the center of mass,  which arises from the repulsive interaction between two nucleons at short $NN$ distances. We note that this depression disappears by kinematical reason when another $N$ ($K^-ppn$) or $\bar{K}$ ($K^-K^-pp$) is added to the $\bar{K}NN$ system.

Figure 4 (right) shows the same distributions but multiplied by $r^2$ ($r^2 \, \rho_{N}(r)$). From these we can see that, when the $\bar{K}N$ attraction is strengthened, strong shrinkage occurs due to an increase of the maxima at around $r = 0.5$ fm and their inward shifts. Especially, the trend of the increase of the maxima in $K^{-}K^{-}pp$ is significant. Besides, from the figure on the left bottom, we can see a growth of the inner amplitude ($r < 0.2$ fm); it remains small for $s_{\bar{K}N}^{(I=0)}$ = -1.2, -1.3, -1.4 and -1.5, but then, it drastically increases for $s_{\bar{K}N}^{(I=0)}$ = -1.6 and -1.7, both in amplitude and in size of this region, showing a growth of a nuclear high-density region in the center of this double-$\bar{K}$ nuclear system.

\subsubsection{$NN$ distance distributions and effective density}

Figure~\ref{fig:Fig4} (left) shows the probability density distributions of the $N$-$N$ distance $\rho_{NN}(r)$ in (a) $K^{-}pp$ (b) $K^{-}ppn$ and (c) $K^{-}K^{-}pp$ for $s_{\bar{K}N}^{(I=0)}$  =  -1.2, -1.3, -1.4, -1.5, -1.6, and -1.7, where $r$ is the $N$-$N$ distance.
We can see that in $K^{-}pp$ and $K^{-}ppn$ $\rho_{NN}(r)$ has a nodal behavior at around $r=0.2$ fm, and has 
an inner amplitude at $r<0.2$ fm. This nodal behavior arises from the rank-2 nature of the $NN$ potential. In $K^{-}K^{-}pp$, this inner amplitude becomes significantly enhanced. The three figures on the right of Fig.~\ref{fig:Fig4} show the same distribution, but multiplied by $r^2$, $r^2 \rho_{NN}(r)$, which is normalized as $\int_{0}^{\infty}dr \, r^2 \rho_{NN}(r)=1$. From these figures, we can see that while the inner amplitude remains small when the $\bar{K}N$ interaction is strengthened in $K^{-}pp$ and $K^{-}ppn$, it becomes much enhanced, and the size of this region also increases (up to around $r=0.5$ fm for $s^{(I=0)}_{\bar{K}N}=$ -1.6 and -1.7) in $K^{-}K^{-}pp$. This indicates the fact that addition of one anti-kaon to $K^{-}pp$ leads to a situation in which two nucleons can get very close to each other in spite of the strong inner repulsion of the $NN$ interaction. 

The $^{1}S_{0}$ $NN$ potential of Ikeda and Sato \cite{Ikeda07}, which we employed in our $K^{-}K^{-}pp$ calculation, reproduces 
the experimental $^{1}S_{0}$ phase shift up to $E_{lab}=300MeV$. This means that the $NN$ interaction is probed up to around $r=0.5$ fm by this potential model. Though the validity of this $NN$ potential for $r<0.5$ fm is not supported, it can be a general argument that the relative motion of two nucleons in the point nucleon model has a non-zero amplitude at $r<0.5$ fm. The present calculation shows that, if there is non-zero amplitude at $r<0.5$ fm, even if it is small, the inner amplitude can grow by coupling to the short-ranged $\bar{K}N$ attraction. If there are two anti-kaons present, the two-fold attraction mechanism leads to a significant growth of this inner amplitude in magnitude and in range, when the $\bar{K}N$ interaction is strengthened.

We calculated the mean $N$-$N$ distances for $K^-pp$, $K^-ppn$ and $K^-K^-pp$, 
\be
R_{NN} = \sqrt{\int _0^\infty  r^4  \rho_{NN}(r) {\rm d} r},
\ee
 for various values of $s_{\bar{K}N}^{(I=0)}$, as shown in Table \ref{tab:table2}. Using the known $NN$ distance \cite{Yamazaki07a,Bethe71},
\be
R_{NN}^{\rm (ord)} = 2.2~{\rm fm},
\ee
for the ordinary nuclear density of $\rho_0 = 0.17$ fm$^{-3}$, we could calculate the ratio, which can be called the ``effective nuclear density" of KNC, as follows: 
\be
\rho^{\rm (eff)} = \left [ \frac{R_{NN}^{\rm (ord)}}{R_{NN}} \right ] ^3 \, \rho_0.
\ee
These values are presented in Table \ref{tab:table2}. For instance, $\rho^{\rm (eff)} = 2.5 \, \rho_0$ in $K^-pp$, and $\rho^{\rm (eff)} = 4.3 \, \rho_0$ in $K^-K^-pp$ at the standard $\bar{K}N$ interaction. The $\rho^{\rm (eff)}$ increases with $|s_{\bar{K}N}^{(I=0)}|$.

\subsubsection{$\bar{K}$-$\bar{K}$ distance distributions}

 To investigate the effect of the repulsion between two $\bar{K}$'s, we show in Fig.~\ref{fig:Fig5} the probability density distribution of the $\bar{K}\bar{K}$ distance 
$\rho_{K^{-}K^{-}}(r)$ (left) and $r^{2}\rho_{K^{-}K^{-}}(r)$ (right) in $K^{-}K^{-}p$ (upper) and $K^{-}K^{-}pp$ (lower).
We note that the $\bar{K}\bar{K}N$ system has an appreciable amplitude for those configurations where the two anti-kaons are very near to each other, already for lower values of $|s_{\bar{K}N}^{(I=0)}|$. By the addition of one nucleon, this inner amplitude grows significantly, and the two $\bar{K}$'s become much closer to each other. Further, the inner amplitude increases drastically when $|s_{\bar{K}N}^{(I=0)}|$ is increased. These features strongly indicate that the effect of the repulsion between the two anti-kaons against the bindings of the $\bar{K}$ nuclear systems is not important. This is against the prevailing belief that because of the $\bar{K}$-$\bar{K}$ repulsion no formation of dense $\bar{K}$ nuclear matter is possible, but is consistent with a dynamical  mechanism for reducing the $\bar{K}$-$\bar{K}$ repulsion due to the $\bar{K} N$ formation \cite{Yamazaki11,Hassanvand11}.

\subsubsection{$\Lambda^*$-$\Lambda^*$ distance distributions in \KKNN}

 Because one $K^-$ and one proton form a compact system ($\Lambda^{*}$), $K^{-}K^{-}pp$ is an important system to study the relative motion of 
two $\Lambda^{*}$ particles. Figure \ref{fig:Fig6} shows the probability density distribution of the distance between the centers of mass of two $\Lambda^{*}$ 
clusters in $K^{-}K^{-}pp$, when $s^{(I=0)}_{\bar{K}N}$ is varied from -1.2 to -1.7. It is interesting that as the attraction between $K^{-}$ and $p$ 
becomes stronger, the maximum of the distribution seems to converge to a distance of around 0.7 fm. In addition, there seems to grow an inner bump 
in the distribution at $r \le 0.3$ fm, particularly  for $s_{\bar{K}N}^{(I=0)} = -1.6$ and $-1.7$. Combined with the strongly enhanced $NN$ inner amplitude for these 
$s_{\bar{K}N}^{(I=0)}$ values, this indicates the possibility of the growth of a very high-density configuration in the center of this double $K^{-}$ 
 nucleus, where all of the two $\bar{K}$'s and two protons become very near to each other.\\

\section{Relativistic effect on $\bar K$ binding}\label{sec:RC}

The calculations so far made are based on a non-relativistic (NR) treatment of few-body systems. For the very deeply bound $\bar K$, however, relativistic corrections are indispensable \cite{Akaishi05}. The relativistic effect can be taken into account by using the Klein-Gordon (KG) equation for $K^-$,
\begin{eqnarray} 
&&\left\{ - \frac{\hbar^2}{2m_K} \vec{\nabla}^2 +U^{\rm opt} \right\}|\Phi \rangle = \left( \epsilon_{\rm KG}+\frac{\epsilon_{\rm KG}^2}{2m_Kc^2} \right) |\Phi \rangle, \nonumber \\ 
&&U^{\rm opt} = U_{\rm s}+U_{\rm v} + \frac{U_{\rm s}^2-U_{\rm v}^2+\epsilon_{\rm KG}U_{\rm v}}{2m_Kc^2},
\end{eqnarray}
where $\epsilon_{\rm KG}$ is the rest-mass-subtracted energy of $K^-$, and $U_{\rm s}$ ($U_{\rm v}$) is a scalar (vector) mean-field potential for $K^-$ from the {\it shrunk nuclear core}. Detailed realistic calculations of $U^{\rm opt}$ are in progress.

If we obtain from few-body NR calculations the lowest energy, $\epsilon_{\rm S}$, of the Schr\"odinger equation with the same optical potential, $U^{\rm opt}$,  
\begin{equation} 
\left\{ - \frac{\hbar^2}{2m_K} \vec{\nabla}^2 +U^{\rm opt} \right\}|\Phi \rangle = \epsilon_{\rm S} |\Phi \rangle, \\ 
\end{equation}
the KG energy is calculated by the relation,
\begin{equation}
\epsilon_{\rm KG} = m_Kc^2 \left(\sqrt{1+\frac{2\epsilon_{\rm S}}{m_Kc^2}}-1 \right). \\ 
\end{equation}
Obviously, this relation means that, when the NR energy, $\epsilon_{\rm S}$, drops down to $-m_Kc^2/2$, the relativistic energy, $\epsilon_{\rm KG}$, becomes $-m_Kc^2$, namely, the total mass of $K^-$ becomes zero ("kaon condensation regime"), as shown in Fig.~\ref{fig:relativity}. This relativistic effect would become particularly large in the case of $K^-K^-pp$, where the core nucleus is strongly shrunk with a highly excited internal energy and, therefore, $| \epsilon_{\rm S} |$ is much bigger than the value naively estimated from $B(K^-K^-pp)$ denoted in Fig 2. \\

Roughly speaking, the relativistic correction for $\epsilon_{\rm S} \sim 100$ MeV is: $\Delta \epsilon_{\rm KG} = \epsilon_{\rm KG} -\epsilon_{\rm S} \approx -10$ MeV. The relativistic correction for $K^-K^-pp$ becomes very large, when the chiral symmetry restoration effect is taken into account.

\section{Effect of chiral symmetry restoration in $\bar{K}$ nuclear systems}\label{sec:chiral}
 
The binding energy of $K^-pp$, $B_K$ = 103 MeV, as observed in the DISTO experiment \cite{Yamazaki10}, turned out to be much larger than the original predicted value (48 MeV \cite{Yamazaki02}, 51.5 MeV in the present work) based on the standard $\Lambda(1405)$ ansatz. This infers that the $I=0$ $\bar{K} N$ interaction is effectively enhanced by about 25\% for some reasons \cite{Yamazaki07b}. 
We examined the problem of the interaction ranges in the adopted $NN$ and $\bar{K}N$ interactions, and found that possible changes of the ranges that are consistent with the mass and width of $\Lambda(1405)$ cannot account for the large binding energy. Thus, the present standard value of the $\bar{K}N$ interaction that corresponds to $B_K$ = 27 MeV of $\Lambda(1405)$ is rather robust; it cannot be increased by 25\%. It is thus plausible to investigate the problem as to  ``how the $\bar{K}N$ interaction changes in the nuclear medium". The relativistic correction in KNC is substantial as we have seen in the preceding Section \ref{sec:RC}, but for $B_K$ = 103 MeV, it is about 10 \% effect, and thus it is not the main cause.

We now consider possible in-medium effects on chiral-symmetry breaking and restoration that occur spontaneously in the meson-nucleon interactions \cite{NJL}. Both the $T$-matrices of the s-wave $\pi N$ and $\bar{K}N$ interactions of Nambu-Goldstone bosons are expressed in relation to the pion and kaon decay constants, $f_{\pi}$ and $f_{K}$, respectively, through the Weinberg-Tomozawa theorem \cite{Weinberg66,Tomozawa66}, 
\begin{eqnarray}
&&T_{\pi N}^{(I=3/2)} = \frac{\omega_{\pi}}{f_{\pi}^2},\\
&&T_{\bar{K}N}^{(I=0)} = - \frac{\omega_K}{f_K^2},
\end{eqnarray}
where the $T$ matrices are inversely proportional to $f_{\pi}^2$ and $f_K^2$. They are further related to the current quark masses, $(m_u, m_d, m_s)$, and quark condensates in the QCD vacuum, according to the Gell-Mann-Oakes-Renner (GOR) relation \cite{GOR}, 
\begin{eqnarray}
m_{\pi}^2 f_{\pi}^2 &=& -4 m_q <0|\bar{q} q|0>, \label{eq:GOR-pi}\\
m_{K}^2 f_{K}^2 &=& - (m_q+m_s) \nonumber\\
  &&\times [<0|\bar{q} q|0> + <0|\bar{s} s|0>], \label{eq:GOR-K}
\end{eqnarray}
where $m_q = (m_u + m_d)/2$, and $<0|\bar{q}q|0> = <0|(\bar{u}u + \bar{d}d)/2|0>$ is a $(u,d)$ quark condensate in free space. The quark condensate is the order parameter of the QCD vacuum \cite{NJL}, and they are subject to changes in the nuclear medium \cite{Hatsuda-Kunihiro85,Hatsuda-Kunihiro94,Waas96,Kaiser95} as $<0|\bar{q} q|0>_{\rm free} \rightarrow <0|\bar{q} q|0>_{\rho}$ in the case of $\pi N$ and $[<0|\bar{q} q|0>_{\rm free} + <0|\bar{s} s|0>_{\rm free}] \rightarrow [<0|\bar{q} q|0>_{\rho} + <0|\bar{s} s|0>_{\rm \rho}]$  in the case of $\bar{K} N$. Our concern is only on the qualitative change of the $T$ matrices, and we expect the following enhancement factors at a nuclear density of $\rho$. For the case of $\pi N$, the ratio of the s-wave isovector interaction with a potential parameter $b_1$ is given by 

\begin{equation}
F_{\pi N} (\rho) = \frac{b_1^{\rho}}{b_1^{\rm free}} = \frac{|<0|\bar{q} q|0>_{\rm free}|}{|<0|\bar{q} q|0>_{\rho}|} \approx  \frac{1}{1 - \alpha \rho},
\label{eq:FpiN}
\end{equation}
with $\alpha$ as a parameter.
Such an enhancement was observed in deeply bound pionic states \cite{Suzuki04,Kienle04,Yamazaki:PhysRep}, where the parameter $\alpha$ was deduced 
  to be $(0.36 \pm 0.08) \times \rho_0^{-1}$ with $\rho_0 = 0.17~ {\rm fm}^{-3}$ being the normal nuclear density. This value is in good agreement with theoretical predictions \cite{Hayano-Hatsuda}.

 In the $\bar{K}N$ sector the enhancement of the $\bar{K}N$ interaction is given by

\begin{equation}
F_{\bar{K} N} (\rho) = \frac{T_{\bar{K}N}^{\rho}}{T_{\bar{K}N}^{\rm free}} = \frac{|<0|\bar{q} q|0>_{\rm free} + <0|\bar{s} s|0>_{\rm free}|}{|<0|\bar{q} q|0> _{\rho}+ <0|\bar{s} s|0>_{\rm \rho}|},
\label{eq:FKbarN}
\end{equation}  
where the $<0|\bar{s} s|0>_{\rm free}$ is known to be around $0.8  <0| \bar{q} q|0>_{\rm free}$ \cite{Reinders85}. 
A similar decrease of the quark condensates, thus, increase of the attractive $\bar{K}N$ interaction, is expected. Here, we can foresee that its effect should be much more spectacular compared with the $\pi N$ case, since the strongly attractive $\bar{K}N$ interaction tends to induce tremendous higher-order effects, in contrast to the case of the $\pi N$ interaction of repulsive character. The most important effect in the $\bar{K}N$ case is substantial nuclear shrinkage caused by the strongly attractive force, as we have considered in the preceding section. So far, no such effect has been taken into account in terms of the modification of the $\bar{K}N$ interaction.  
  
In the present note we attempt to understand this effect qualitatively by applying the concept of ``clearing of QCD vacuum" introduced by Brown, Kubodera and Rho \cite{BKR}. This model states that the QCD vacuum expressed by the quark condensate is reduced by the amount of clearing of the QCD vacuum by the presence of nucleons. 
 We extend the BKR model to $\bar{K}$ nuclear bound systems (with a nucleon number $n$), postulating the following enhancement factors for the $\pi N$ and $\bar{K}N$ cases, respectively:
\bea
&&F_{\pi N} = \frac{|<0|\bar{q} q|0>_{0}|}{|<0|\bar{q} q|0>_{\rho}|} \approx  \frac{1}{1 - \Omega}, \label{eq:condensate-pi}\\
&&F_{\bar{K}N} = \frac{|<0|\bar{q} q|0>_{0}|}{|<0|\bar{q} q|0>_{\rho}|} \approx  \frac{1}{1 - 0.5\,\Omega}
\label{eq:condensate-K}\eea 
with a ``QCD vacuum clearing" factor of $\Omega$, which is given by the ratio of the volume of the nucleon ($v_N$) that clears the QCD vacuum to the nuclear volume ($V_{\rm nucl}$) as
\begin{equation}
\Omega = \frac{n \cdot v_N}{V_{\rm nucl}}.\label{eq:Omega}
\end{equation}
We take an ``$rms$ volume''  for $v_N$,
\begin{equation}
v_N = \frac{4 \pi}{3} {r_N}^3 = 2.85 ~[{\rm fm}^3],
\end{equation}
corresponding to an {\it rms} nucleon radius of $r_N = 0.88$ fm \cite{PDG}, since it represents a dense region of the nucleon that is expected to clear the volume of the QCD vacuum. In the case of the $\bar{K}N$ case, the denominator in eq.(\ref{eq:condensate-K}) is approximated to $1 - 0.5 \, \Omega$, since the chiral restoration effect dominates the $<0|\bar{q} q|0>$ term in eq.(\ref{eq:FKbarN}), so that $<0|\bar{s} s|0>_{\rho}$ is close to $<0|\bar{s} s|0>_{\rm free}$ \cite{Hatsuda-Kunihiro94}. 

In the case of a normal nucleus having a mass number $A$ and a nuclear radius of $R_A = 1.2 \times A^{1/3}$ fm at the standard nuclear density $\rho_0 = 0.17$ fm$^{-3}$, the nuclear volume is 
\begin{equation}
V_{\rm nucl} = 7.24  \times A ~ [{\rm fm}^3].
\label{eq:Vnucl} 
\end{equation}
Thus, 
\begin{equation}
\Omega \approx 0.34\, \frac{\rho}{\rho_0}, 
\end{equation}
which gives good agreement with the observation of $F_{\pi N} (\rho)$ in the deeply bound pionic case, eq. (\ref{eq:FpiN}).

In the case of $\bar{K}$ clusters, how to define $V_{\rm nucl}$ is not obvious. Since  few-body $\bar{K}N$ clusters do not provide a homogeneous medium, we evaluated $V_{\rm nucl}$ from $V_{\rm KNC}$ using the realistic wave functions. Figure~\ref{fig:dango} shows the nucleon distributions $\rho_{N}(r)$ and $4 \pi r^2 \rho_{N}(r)$ for $K^-pp$ and $K^-K^-pp$ both for point and finite-size nucleons in a schematical figure. We can define an effective radius, $R_{\rm eff}$, for ``80 \% volume" as
\be
\int_{0}^{R_{\rm eff}} r^2 \rho_N^{\rm finite} (r) \, {\rm d}r = n \times 0.80.
\ee
The effective nuclear volume is 
\be
V_{\rm eff} = \frac {4 \pi}{3} {R_{\rm eff}}^3,
\ee
which yields $\Omega_{\bar{K}N} = n \cdot v_N/V_{\rm eff}$, as shown in Table~\ref{tab:table3}. Using these values we can obtain enhancement factors for the $\bar{K} N$ interactions, $F_{\bar{K}N}$, which lead to enhanced interaction strengths, $s_{\bar{K}N}^{(1st)}$, with the same non-dimensional unit as $s_{\bar{K}N}$, where the suffix $(1st)$ means the first iteration of such renormalization.

\begin{table*}
\begin{center}
\caption{\label{tab:table3}  
Effect of chiral symmetry restoration in $K^{-}pp$, $K^{-}ppn$ and $K^{-}K^{-}pp$. 
$R_{\rm eff}$ is the radius corresponding to 80 \% volume of each $KNC$: $V_{\rm eff}$ [fm$^3$]. $\Omega_{\bar{K}N}$ = QCD-vacuum clearing factor. 
$F_{\bar{K}N}$ = enhancement factor of the $\bar{K}N$ interaction. $s_{\bar{K}N}^{\rm (1st)}$ = renormalized interaction strength after the 1st iteration.}

\vspace{0.5cm}
\begin{tabular}{l|ccccc|ccccc|ccccc}
\hline
 &   \multicolumn{5}{c}{$K^-pp$} & \multicolumn{5}{c}{$K^-ppn$} & \multicolumn{5}{c}{$K^-K^-pp$} \\     
 $s_{\bar{K}N}$ &   $R_{\rm eff}$ &  $V_{\rm eff}$  & $\Omega_{\bar{K}N}$ & $F_{\bar{K}N}$~& $s_{\bar{K}N}^{(\rm 1st)}$ ~
                 &   $R_{\rm eff}$ &  $V_{\rm eff}$  & $\Omega_{\bar{K}N}$ & $F_{\bar{K}N}$ & $s_{\bar{K}N}^{(\rm 1st)}$ 
                 &   $R_{\rm eff}$ &  $V_{\rm eff}$  & $\Omega_{\bar{K}N}$ & $F_{\bar{K}N}$ & $s_{\bar{K}N}^{(\rm 1st)}$\\
 \hline
$-1.2$ &   1.68 & 19.8 & 0.29 & 1.17  & -1.40         & 1.77 & 23.1 & 0.37 & 1.23 & -1.47   &   1.55 & 15.6 & 0.37  & 1.22 &   -1.47 \\ 
$-1.3$ &  1.58 & 16.5  & 0.35 & 1.21  & -1.57        & 1.71 & 21.0 & 0.41 & 1.26 & -1.63    &  1.47 & 13.4 & 0.43  & 1.27 &  -1.65 \\
${\bf -1.37}$  & {\bf 1.53} & {\bf 15.0} & {\bf 0.38} & {\bf 1.24}  & {\bf -1.69}  & {\bf 1.68} & {\bf 19.9} & {\bf 0.43} & {\bf 1.27} & {\bf -1.75}   & {\bf 1.43} & {\bf 12.3} & {\bf 0.46}  & {\bf 1.30} &  {\bf -1.78}    \\
$-1.4$ &   1.51 & 14.5   & 0.39 & 1.25  & -1.74          & 1.67 & 19.6 & 0.44 & 1.28 & -1.79  &   1.42 & 11.9 & 0.48  & 1.31 & -1.84  \\
$-1.5$ &   1.46 & 13.1   &  0.43 & 1.28 & -1.92          & 1.63 & 18.3 & 0.47 & 1.31   & -1.96    & 1.38 & 11.0 & 0.52 & 1.35 & -2.02 \\
$-1.6$  & 1.43 & 12.2   &  0.47 & 1.31 & -2.09         & 1.61 & 17.4 & 0.49 & 1.33  & -2.12      & 1.35 & 10.4 & 0.55 & 1.38 & -2.21 \\
$-1.7$   & 1.40 & 11.4  & 0.50  & 1.33 & -2.26          & 1.58 & 16.6 & 0.52 & 1.35 & -2.29    &  1.34 & 10.0 & 0.57  & 1.40 & -2.38  \\
 \hline
\end{tabular}
\end{center}
\end{table*}

 Let us take $K^-pp$ as an example. As the first step, we take $s_{\bar{K}N}^{(I=0)} = -1.37$ so as to be consistent with the assumed binding energy of 27 MeV of $K^-p$ identified as $\Lambda(1405)$. The BKR procedure gives a renormalized interaction parameter, enhanced by a factor of 1.24. This in turn gives a renormalized interaction strength: $s_{\bar{K}N}^{(I=0)} = -1.37 \times 1.24 = - 1.69$. It is known that the experimental binding energy of $K^-pp$, 103 MeV, obtained from the DISTO experiment \cite{Yamazaki10}, is about a factor of 2 larger than the original few-body predictions taking into account the $NN$ repulsion appropriately, 53 MeV \cite{Yamazaki02}, 55-70 MeV \cite{Shevchenko07a}, 76 MeV \cite{Ikeda07}, and 51.5 MeV (present). This discrepancy, as shown in Fig. 2, can now be explained by an increase of the $\bar{K} N$ interactions due to chiral symmetry restoration.

Unfortunately, there is only one empirical data that we can take into account with confidence. 
 We look forward to more experimental data to be produced in the near future.

\section{Concluding remarks}

We have made a comprehensive systematic study for the kaonic nuclear systems $K^{-}p$,
$K^{-}K^{-}p$, $K^{-}pp$, $K^{-}ppn$ and $K^{-}K^{-}pp$ by employing separable
potential models for the $NN$, $\bar{K}N$ and $\bar{K}\bar{K}$ interactions. Our calculations based on the PDG interpretation of
$\Lambda(1405)$ show deep bindings of $K^{-}pp$, $K^{-}ppn$ and $K^{-}%
K^{-}pp$ in contrast to the weak binding results based on the
chiral SU(3) dynamics model of the $\bar{K}N$ interaction.

To globally view the deep nuclear binding and strong shrinkage induced by the
short-ranged strong $\bar{K}N$ interaction, we created a reference
diagram of the binding enegies of these kaonic nuclear systems with respect to
the strength of the $\bar{K}N$ attraction, and calculated the probability
density distributions of the nucleons and the relative distance of the two
nucleons from the obtained wave function. Besides the increase in binding
energies with the increase in the $\bar{K}N$ attraction, there appears to be a 
considerable shrinkage behavior of the size of the nucleonic motion in these
$\bar{K}$ nuclear systems, as can be seen from Fig~\ref{fig:Fig3}. A main shrinkage occurs by inward shifts of the maxima in $r^2 \rho (r)$ at around 0.5 fm and of the tail parts ($r > 1$ fm). 

Especially, we can see a 
notable shrinkage in $K^{-}K^{-}pp$, where the ``super-strong
nuclear force" mediated by an anti-kaon between two nucleons \cite{Yamazaki07a,Yamazaki07b} works twofold, in
a simple counting, compared with $K^{-}pp$. The repulsive force between the two
anti-kaons is not effective to prevent this double attraction mechanism (see also \cite{Yamazaki11,Hassanvand11}). 
Surprisingly, as shown in Fig.~\ref{fig:Fig4} (c), the $NN$ relative wave function reveals  an appreciable inner
amplitude at very short $NN$ distances ($r_{NN} < 0.5$ fm, inside of the size of the nucleons
themselves), whereas  $K^{-}pp$ and $K^{-}ppn$ show only small amplitudes (Fig.~\ref{fig:Fig4} (a,b)). 
Our calculation indicates that this inner amplitude in the $K^{-}
K^{-}pp$ system grows appreciably when we take around $s_{KN}^{(I=0)}=-1.5$ or stronger values. This behavior of the enhanced $NN$ inner amplitudes suggests that the
short-ranged attraction due to the exchange of $\rho$ meson, which is 
taken into account in the range of our $\bar{K}N$ interaction as the underlying process of the Weinberg-Tomozawa term, sensitively
couples to the two nucleons when located almost next to each other, thus generating the growth of extremely
high-density nuclear configurations in the central region of the $\bar{K} \bar{K}$ 
nuclear systems.

In analyzing the DISTO experiment, a compact $K^{-}pp$\ state is
obtained, the binding energy of which is much deeper than the original theoretical
predictions. To clarify this situation, we conducted an analysis on the effect of
the partial restoration of chiral symmetry on the $\bar{K}N$ interaction on
the basis of the ``clearing QCD vacuum" model of Brown, Kubodera and Rho \cite{BKR}. In
this analysis, we calculated the QCD vacuum clearing factor, $\Omega$, from the
obtained nucleon density distributions. Our analysis shows that the compact
$K^{-}pp$\ state indicated from the DISTO experiment is consistent with the
strengthening of the $\bar{K}N$ interaction due to the partial restoration of
the chiral symmetry. Our analysis
further suggests an even stronger $\bar{K}N$ interaction in the $K^{-}K^{-}
pp$\ system, i.e. a very compact $K^{-}K^{-}pp$\ state. This could thus be a
possible mechanism to generate extremely high density, and requires further theoretical studies. For a systematic
understanding of the chiral symmetry restoration in dense nuclear matter, we
strongly encourage efforts to conduct experimental searches for kaonic nuclear bound
systems including $K^{-}K^{-}pp$, in addition to $K^{-}pp$ \cite{Kienle:07,Zmeskal:09,Herrmann:13,Yamazaki:12}.\\

We would like to dedicate the present paper to the late Professor Paul Kienle, who has shown great enthusiasm and stimulation to our work. We would like to thank Dr. Mannque Rho for the illuminating discussion. We acknowledge the receipt of the Grant-in-Aid of Monbu-Kagakusho of Japan.



\begin{figure*}[htb]
\vspace{-0.5cm}
\includegraphics[width=0.7\textwidth]{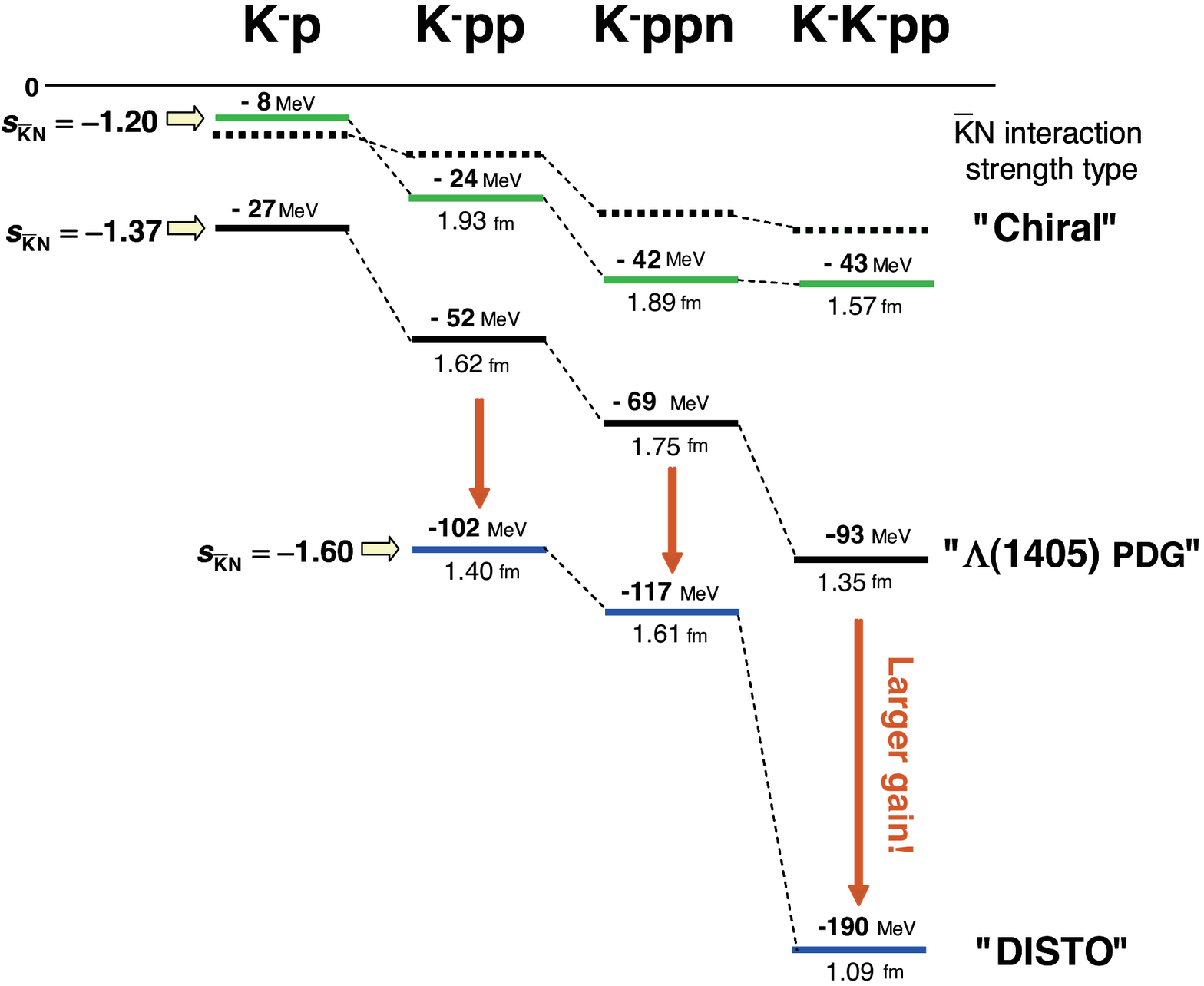}
\caption{\label{fig:Overview} Global view of the calculated bound-state energies $E_B$ (in MeV) and sizes $R_{NN}$ (in fm) of $\bar{K}$ nuclear clusters, $K^-p$, $K^-pp$, $K^-ppn$ and $K^-K^-pp$ in the ``Chiral" model ($s_{\bar{K}N}^{(I=0)} = -1.20$) (green bars) and in the ``Standard" regime ``$\Lambda(1405)$ PDG" ($s_{\bar{K}N}^{(I=0)} = -1.37$) (black bars) and an enhanced regime ``DISTO" ($s_{\bar{K}N}^{(I=0)} = -1.60$) (blue bars). The chiral-based calculations of Barnea {\it et al.} \cite{Barnea12} are shown by black broken bars. }
\end{figure*} 

\begin{figure*}[htb]
\vspace{-0.5cm}
\includegraphics[width=0.7\textwidth]{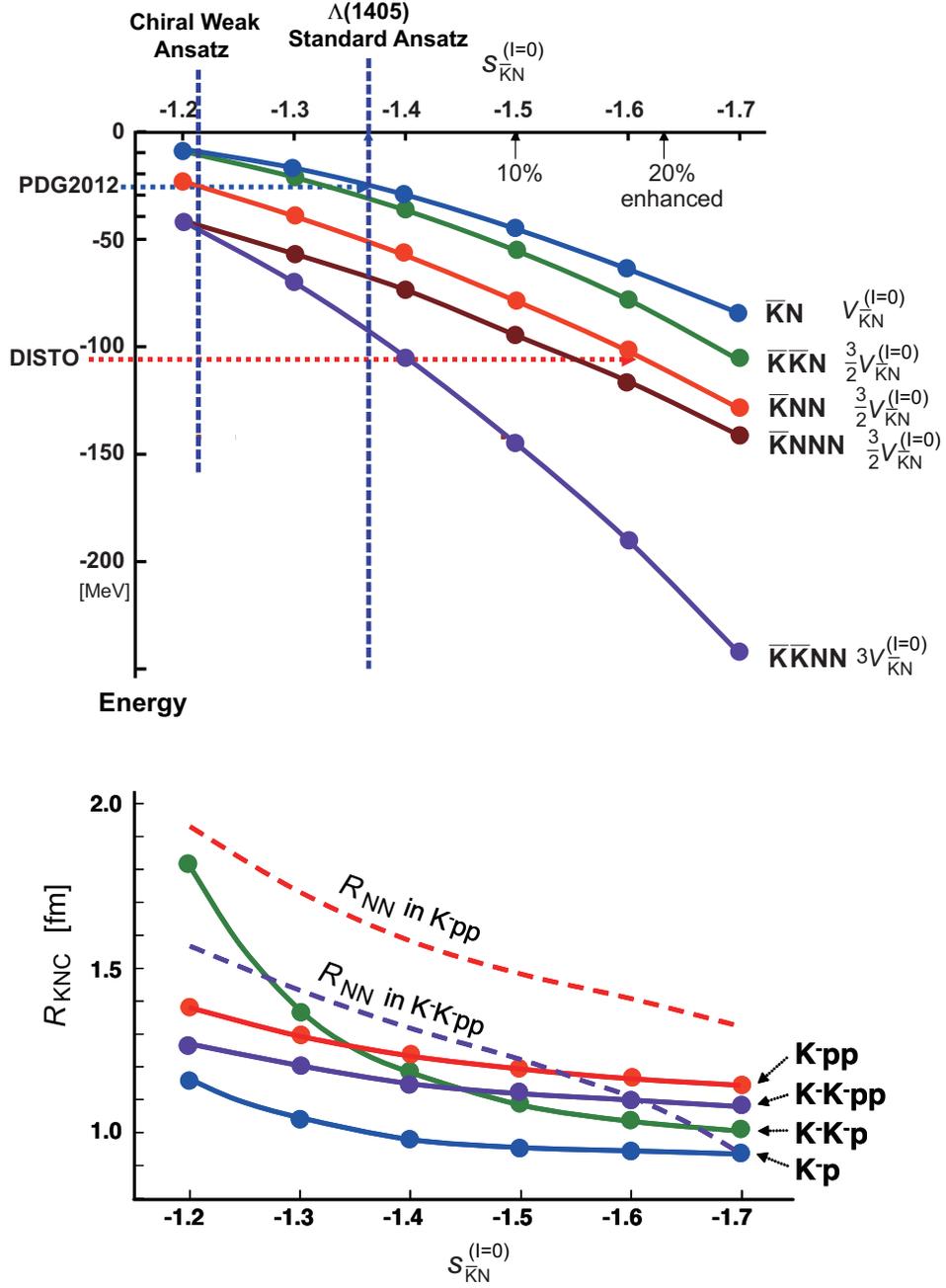}
\caption{\label{fig:Global}Global view of the calculated bound-state energies (upper) and sizes (lower), $R_{\rm KNC}$ and $R_{NN}$ of $\bar{K}$ nuclear clusters as functions of the $\bar{K}N$ interaction strength, $s_{\bar{K}N}^{(I=0)}$. The zones of the standard ``$\Lambda(1405)$ ansatz" and the ``Chiral" ansatz are shown by vertical broken lines. The experimental value of the mass of $K^-pp$ as observed by DISTO \cite{Yamazaki10} 
is shown by a horizontal broken line, where a relativistic correction for the binding energy around 10 MeV is not taken into account.}
\end{figure*} 

\begin{figure*}[htb]
\vspace{0cm}
\includegraphics[width=0.8\textwidth]{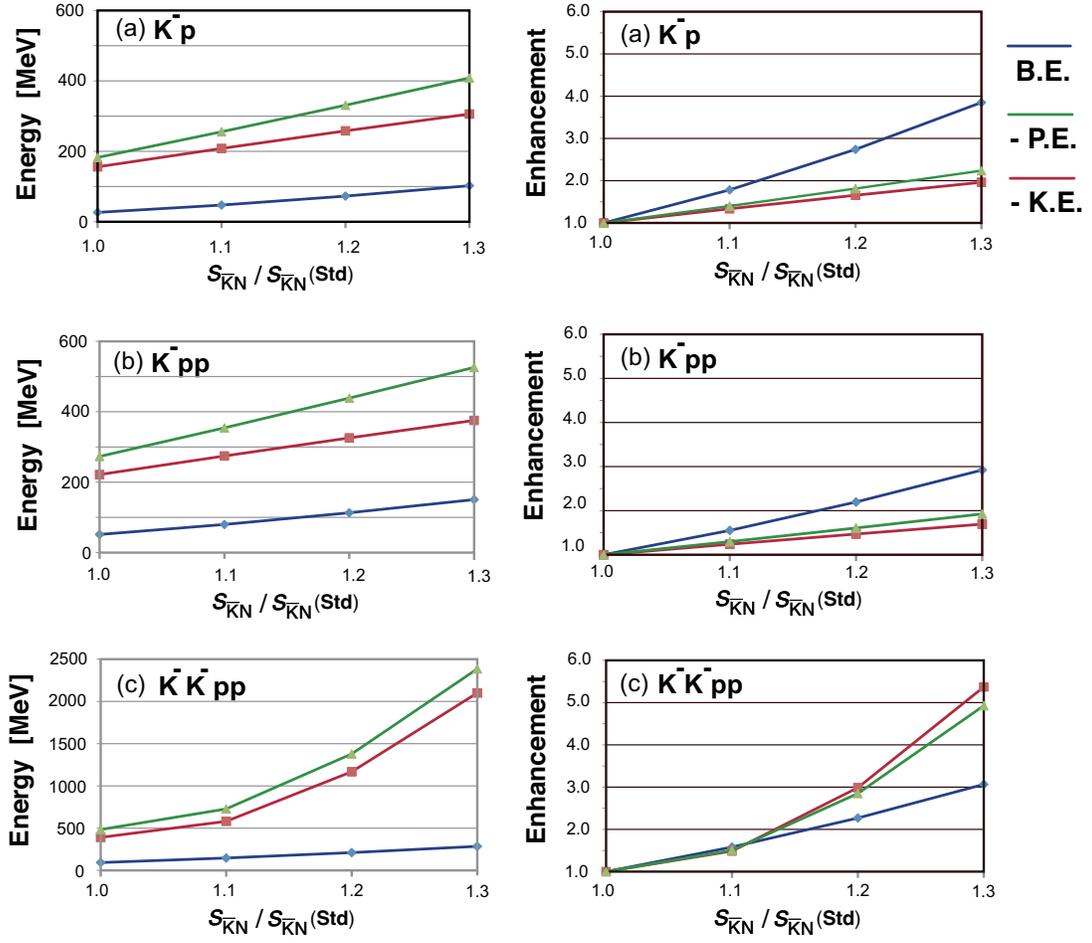}
\caption{\label{fig:Higher-order-effect} Illustration of the energy values of binding (B.E.), potential (P.E.) and kinetic (K.E.) energies (left) and their enhancements (right) as functions of the interaction strength $f = s_{\bar{K}N}^{(I=0)} / s_{\bar{K}N}^{(I=0)} (\rm Std)$, indicating enormous higher-order effects of wave function shrinkage caused by the increase of $\bar{K}N$ attraction.}
\end{figure*} 

\begin{figure*}[htb]
\vspace{-0.5cm}
 \includegraphics[width=0.45\textwidth]{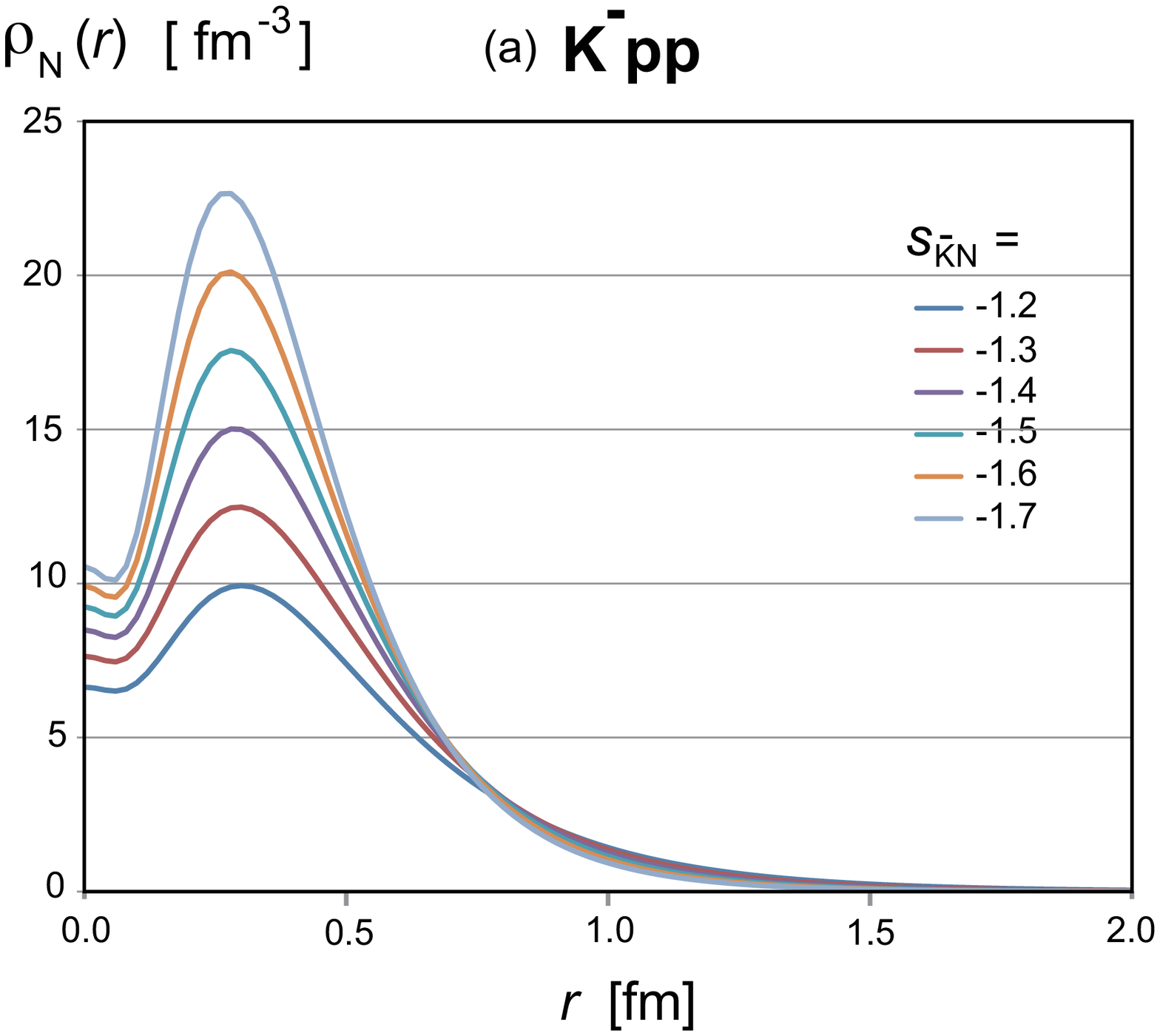}
  \includegraphics[width=0.45\textwidth]{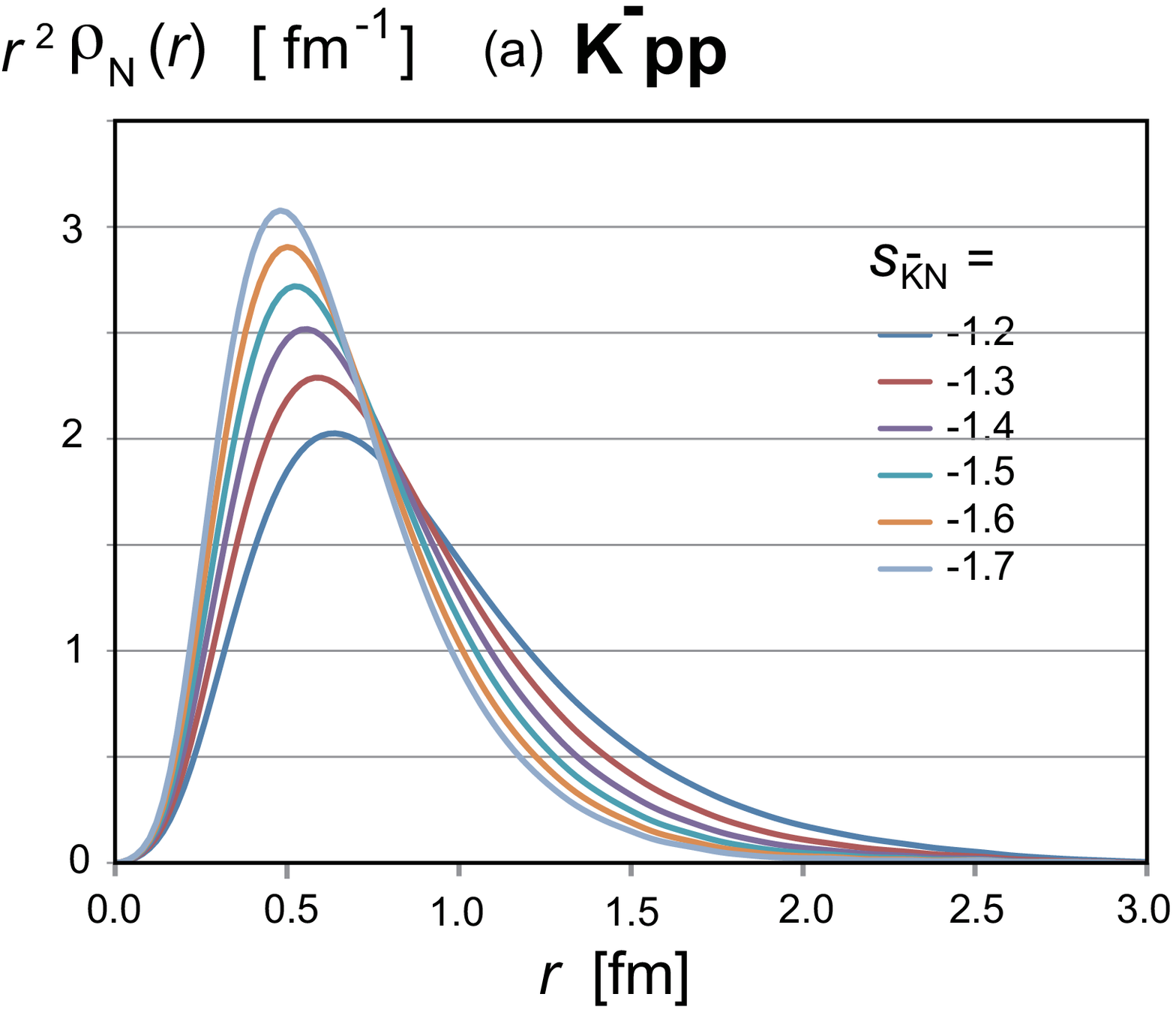}\\
 \includegraphics[width=0.45\textwidth]{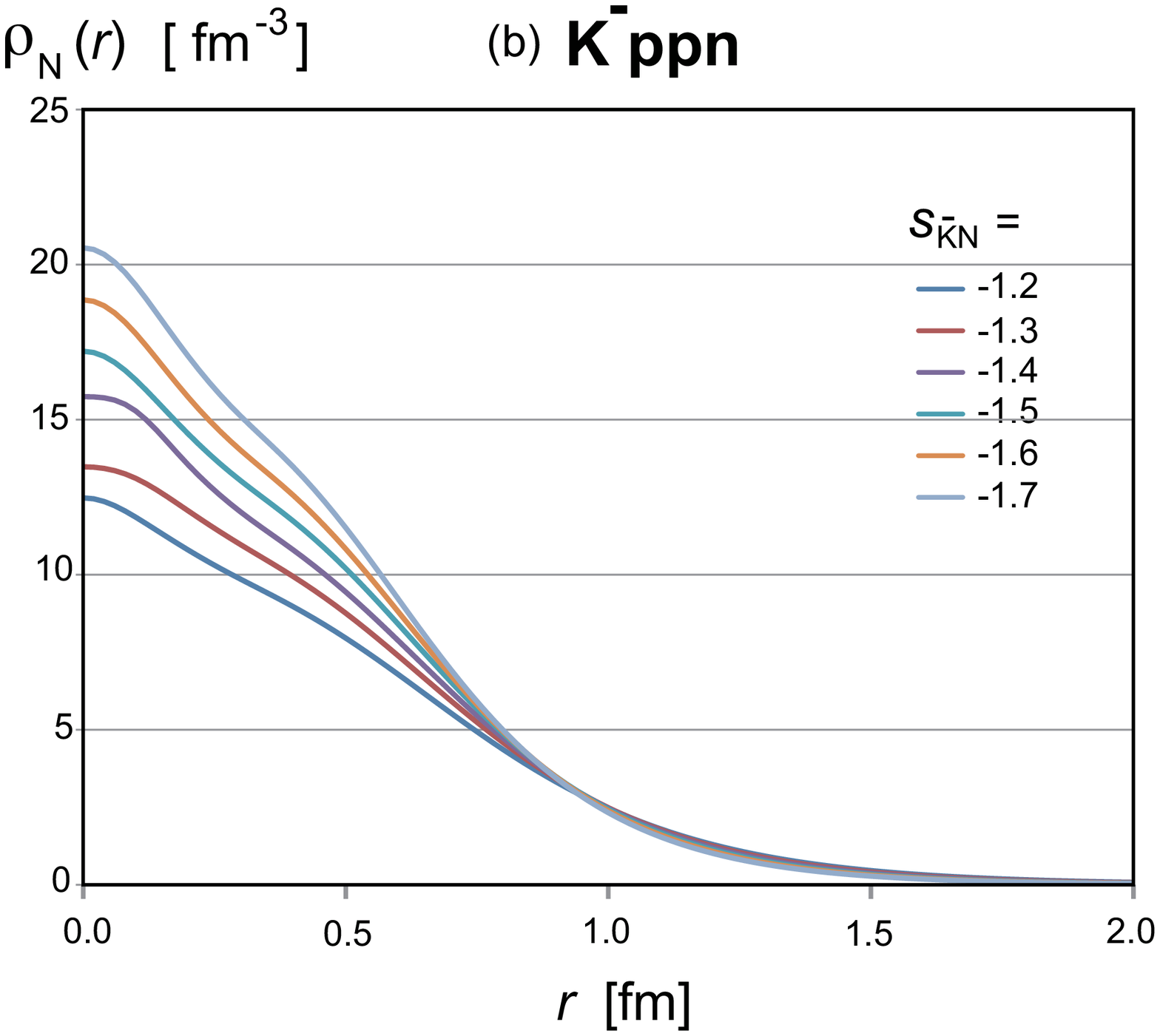}
\includegraphics[width=0.45\textwidth]{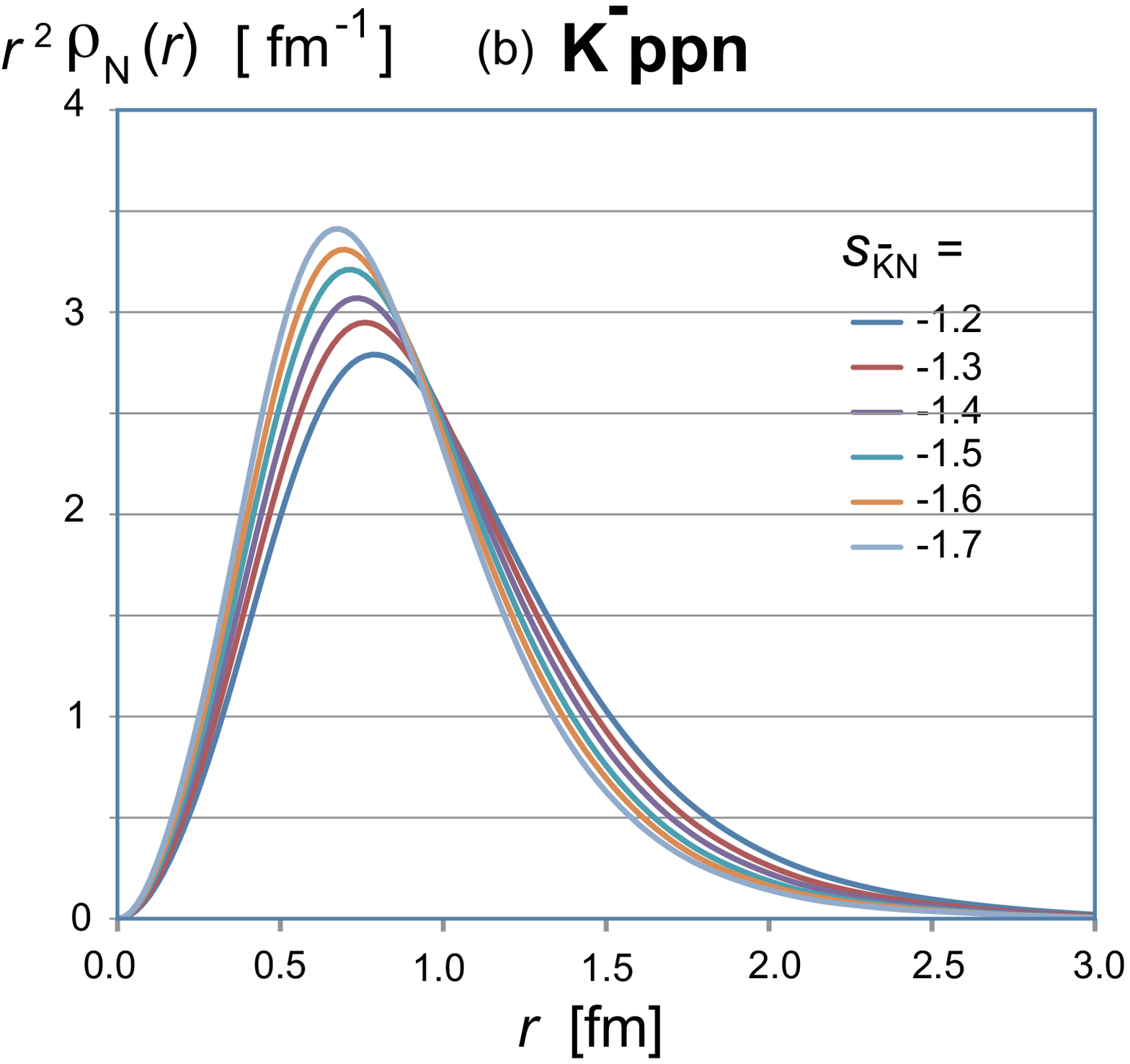}\\
 \includegraphics[width=0.45\textwidth]{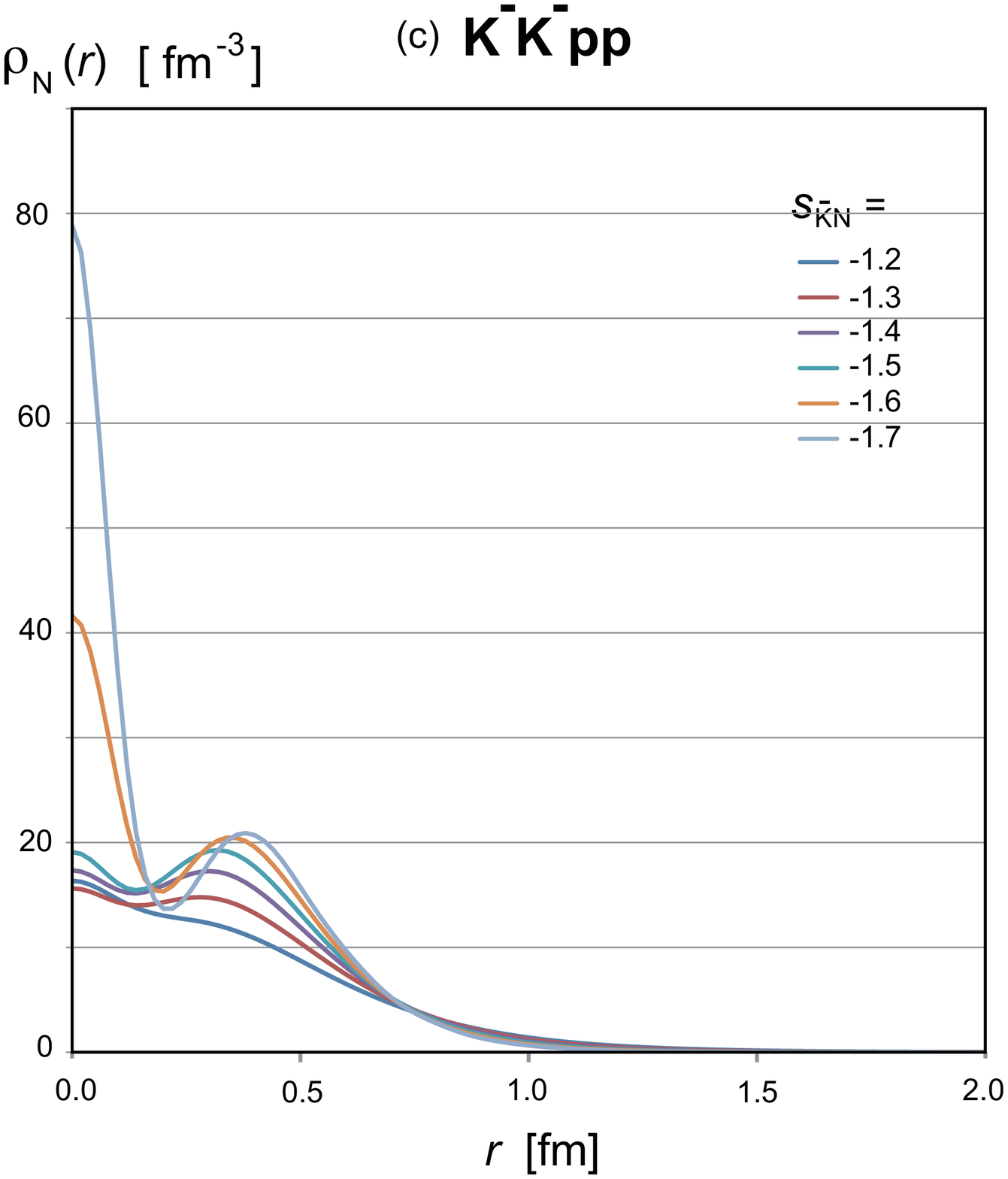}
 \includegraphics[width=0.45\textwidth]{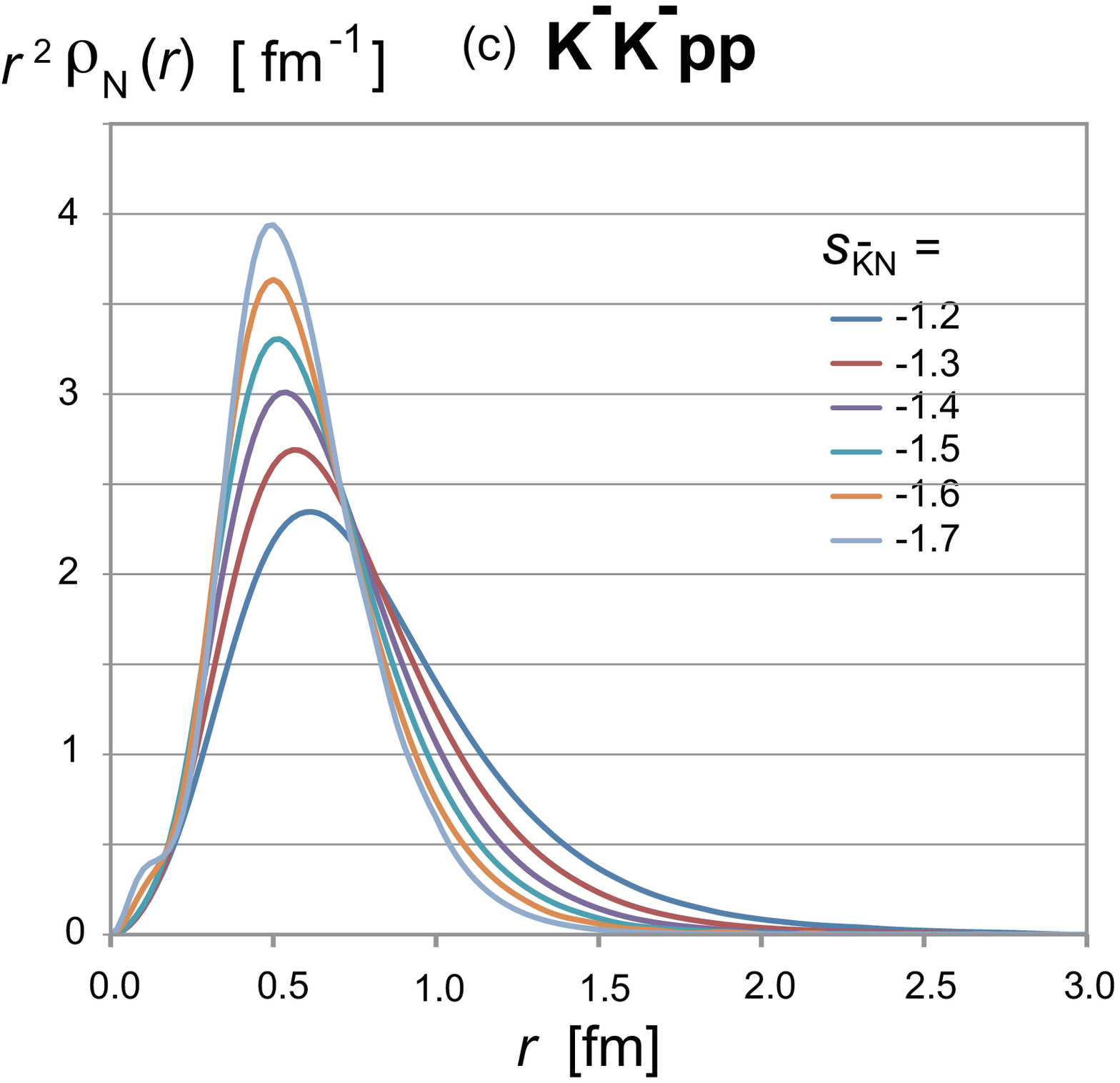}
\caption{\label{fig:Fig3}
Nucleon density distributions, (left) $\rho_{N}(r)$ and (right) $r^2 \rho_{N}(r)$, with respect to the center of mass for (a) $K^{-}pp$, (b) $K^{-}ppn$ and (c) $K^{-}K^{-}pp$ for various values of $s^{(I=0)}_{\bar{K}N}$ = -1.2, -1.3, -1.4, -1.5, -1.6 and -1.7. They are normalized so that $\int_0^\infty r^2 {\rm d}r \rho (r)$ =  number of the nucleons. No nucleon size is included. }
\end{figure*} 

\begin{figure*}[htb]
\vspace{-0.5cm}
\includegraphics[width=0.95\textwidth]{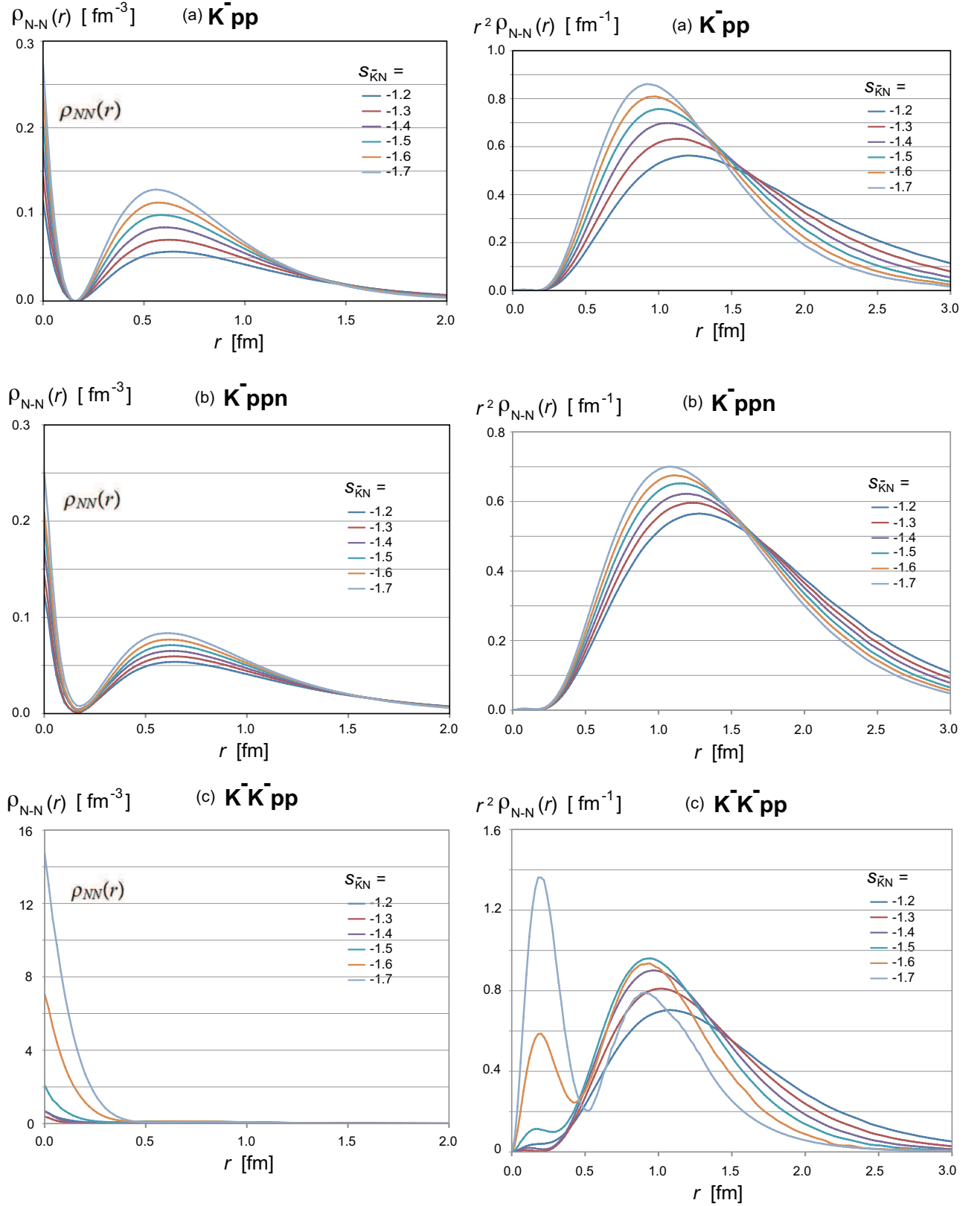}\\
\caption{\label{fig:Fig4} $NN$ distance distributions (left) $\rho_{N-N}(r)$ and (right) $r^{2}\rho_{N-N}(r)$ in (a) $K^-pp$, (b) $K^- ppn$ and (c) $K^-K^- pp$ for $s_{\bar{K}N}^{(I=0)}$ = -1.2, -1.3, -1.4, -1.5, -1.6, and -1.7. No nucleon size is included. As the absolute value of $s_{\bar{K}N}^{(I=0)}$ becomes larger, the distribution in the tail region becomes relatively more pulled inward.}
\end{figure*}

\begin{figure*}[htb]
\vspace{-0.5cm}
\includegraphics[width=0.9\textwidth]{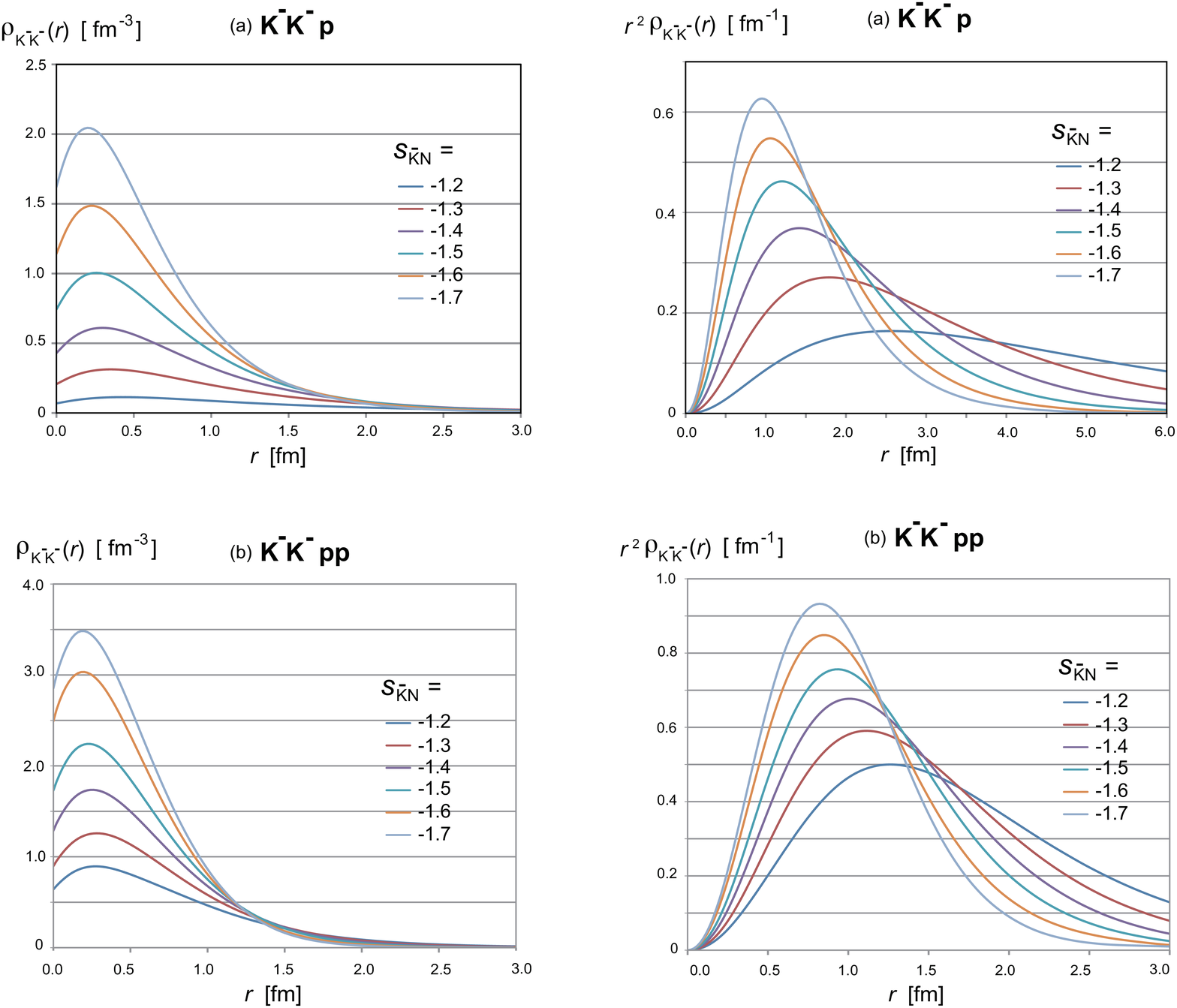}\\
\caption{\label{fig:Fig5}  Probability distribution of the $\bar{K}\bar{K}$ distance (left) $\rho_{K^{-}K^{-}}(r)$ and (right) $r^{2}\rho_{K^{-}K^{-}}(r)$ in (upper) $K^{-}K^{-}p$ and (lower) $K^{-}K^{-}pp$.}
\end{figure*} 

\begin{figure*}[htb]
\vspace{+0.5cm}
\includegraphics[width=0.9\textwidth]{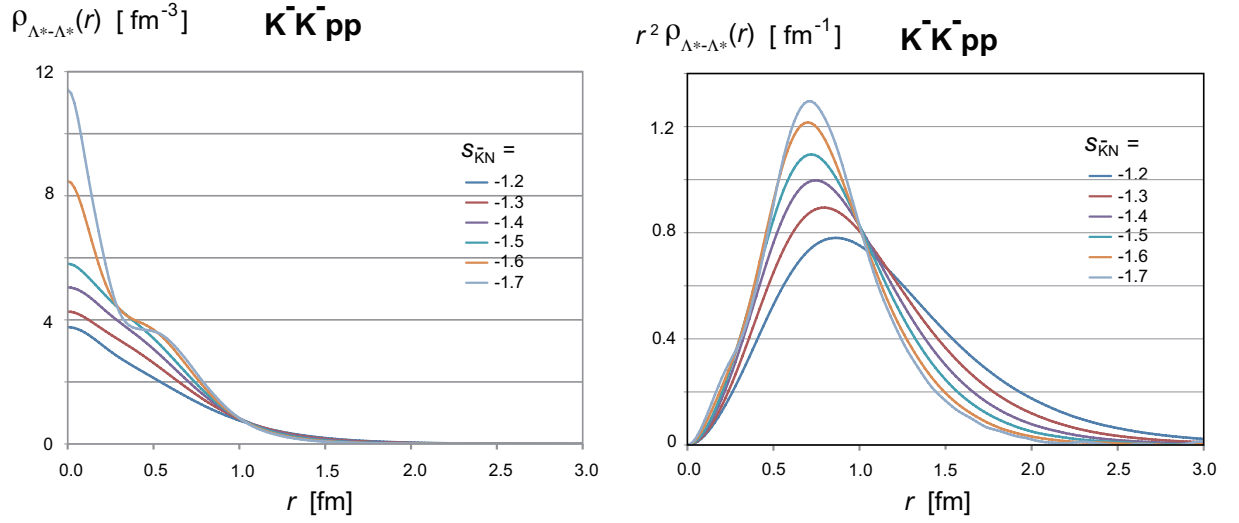}\\%
\caption{\label{fig:Fig6}  
Probability distribution of the distance between the centers of mass of two $\Lambda^{*}$ clusters in $K^{-}K^{-}pp$, 
when $s_{\bar{K}N}^{(I=0)}$ is varied from -1.2 to -1.7.
}
\end{figure*} 

\begin{figure}[htb]
\vspace{+0cm}
\includegraphics[width=7cm]{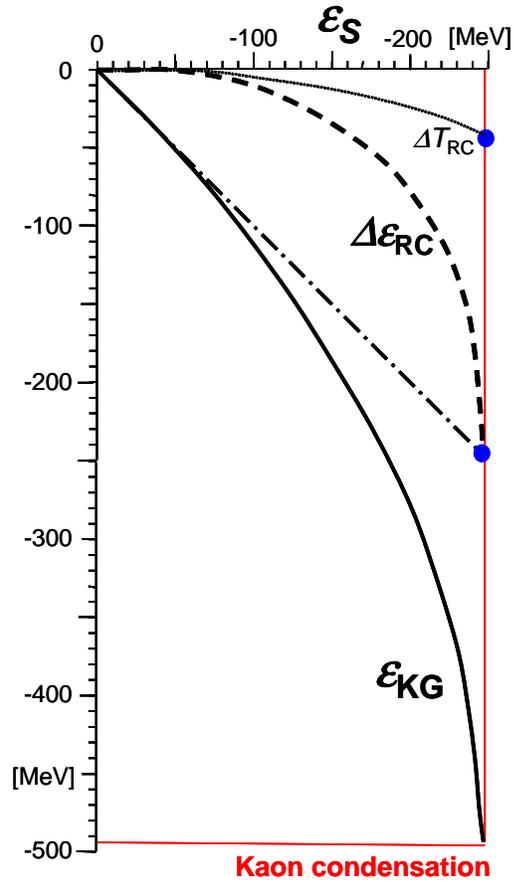}%
\caption{\label{fig:relativity} 
 Comparison between the Klein-Gordon energy, $\epsilon_{\rm KG}$, and the Schr\"odinger energy, $\epsilon_{\rm S}$. The relativistic correction, $\Delta \epsilon_{\rm RC} =  \epsilon_{\rm KG} - \epsilon_{\rm S}$, is shown by a broken curve. As $\epsilon_{\rm S} \rightarrow - m_K c^2/2$, $\epsilon_{\rm KG}$ becomes $-m_K c^2$ ({\it Kaon condensation})}. 
 \end{figure}

\begin{figure*}[htb]
\vspace{+0.5cm}
\includegraphics[width=0.9\textwidth]{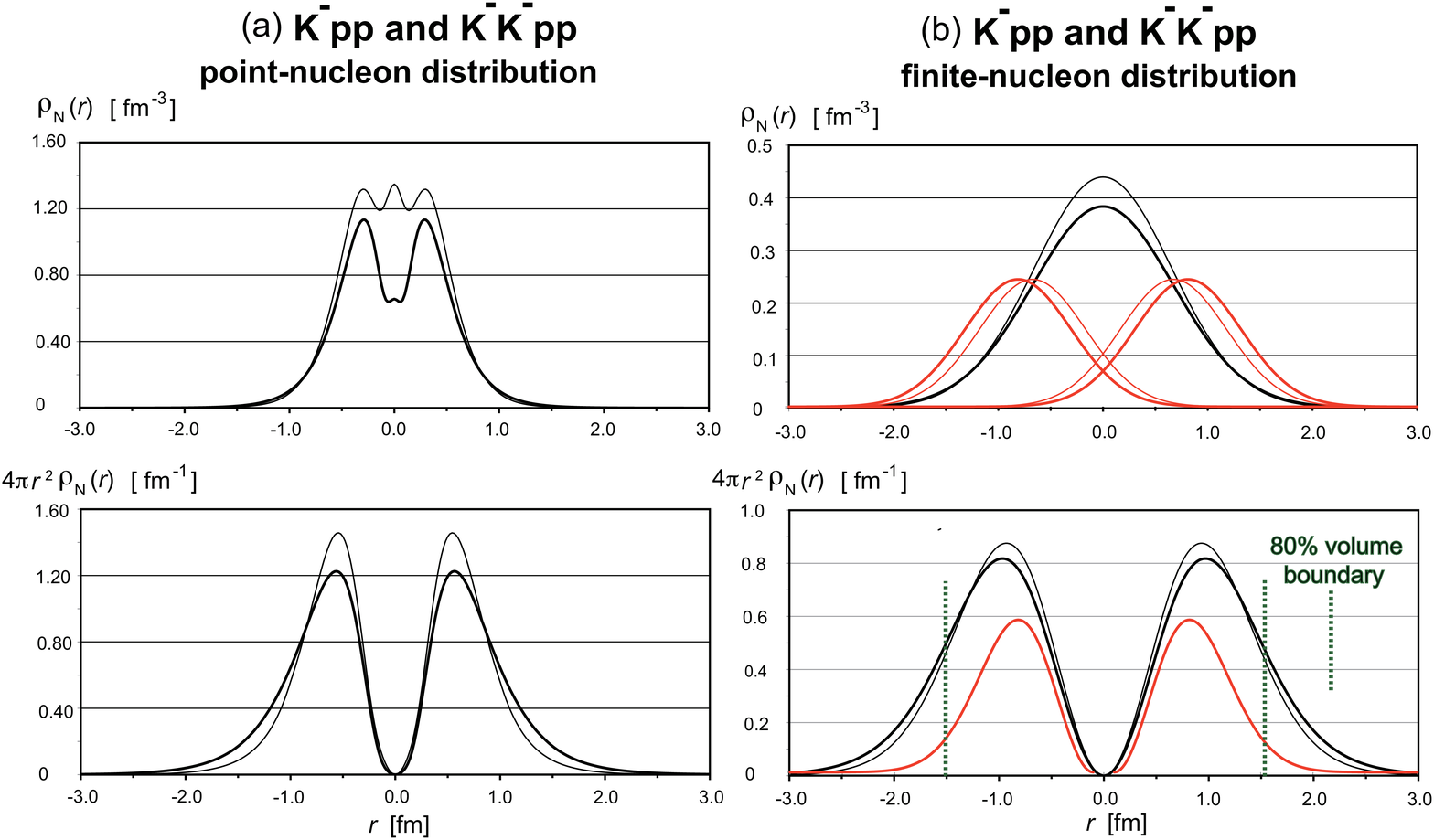}\\%
\caption{\label{fig:dango} 
 @Nucleon density distributions in $K^-pp$ (solid curves) and $K^-K^-pp$ (thin curves) at $s_{\bar{K}N}^{(I=0)}$ (Std) = -1.37, a) for a point nucleon and b) for a finite nucleon, (upper) $\rho_{N}(r)$ and (lower) $4 \pi r^2 \rho_{N}(r)$. The red curves represent the proton finite distributions when the two protons are located at the mean separation distances, $R_{NN}$. }
\end{figure*} 

\end{document}